\documentclass{article}

\usepackage{amssymb}
\usepackage{amsthm}
\usepackage{graphicx}
\usepackage{subfigure}
\usepackage{wrapfig}
\usepackage{xspace}
\usepackage[a4paper]{geometry}

\begin{document}

\title{An annotated bibliography on 1-planarity}

\author{Stephen~G.~Kobourov$^1$, Giuseppe~Liotta$^2$, Fabrizio~Montecchiani$^2$
\\[0.2in]
$^1$Dep. of Computer Science, University of Arizona, Tucson, USA\\\texttt{\small kobourov@cs.arizona.edu} \and
$^2$Dip. di Ingegneria, Universit{\`a} degli Studi di Perugia, Perugia, Italy\\\texttt{\small\{giuseppe.liotta,fabrizio.montecchiani\}@unipg.it}}

\maketitle

\begin{abstract}
The notion of  1-planarity is among the most natural and most studied generalizations of graph planarity. A graph is 1-planar if it has an embedding where each edge is crossed by at most another edge. The study of 1-planar graphs dates back to more than fifty years ago and, recently, it has driven increasing attention in the areas of graph theory, graph algorithms, graph drawing, and computational geometry. This annotated bibliography aims to provide a guiding reference to researchers who want to have an overview of the large body of literature about 1-planar graphs. It reviews the current literature covering various research streams about 1-planarity, such as characterization and recognition, combinatorial properties, and geometric representations. As an additional contribution, we offer a list of open problems on 1-planar graphs.
\end{abstract}

\section{Introduction}

Relational data sets, containing a set of objects and relations between them, are commonly modeled by graphs, with the objects as the vertices and the relations as the edges.
A great deal is known about the structure and properties of special types of graphs, in particular {\em planar graphs}. The class of planar graphs is fundamental for both Graph Theory and Graph Algorithms, and is extensively studied. Many structural properties of planar graphs are known and these properties can be used in the development of efficient algorithms for planar graphs, even where the more general problem is NP-hard~\cite{Baker1994}.

Most real world graphs, however, are {\em non-planar}. For example, scale-free networks (which can be used to model web-graphs, social networks and biological networks) consist of sparse non-planar graphs. To analyze such real world networks, we need to address fundamental mathematical and algorithmic challenges for {\em sparse non-planar} graphs, which we call {\em beyond-planar graphs}. Beyond-planar graphs are more formally defined as non-planar graphs with topological constraints, such as forbidden crossing patterns, as is the case with graphs where the number of crossings per edge or the number of mutually crossing edges is bounded by a constant (see, e.g.,\cite{FoxPS13,PachT97,SukW15}).

A natural motivation for studying beyond-planar graphs is visualization. The goal of graph visualization is to create a good geometric representation of a given abstract graph, by placing all the vertices and routing all the edges, so that the resulting drawing represents the graph well. Good graph drawings are easy to read, understand, and remember; poor drawings can hide important information, and thus may mislead.
Experimental research has established that good visualizations have a number of geometric properties, known as {\em aesthetic criteria}, such as few edge crossings, symmetry, edges with low curve complexity, and large crossing angles; see~\cite{HuangEH14,Purchase97,Purchase00,PurchaseCA02}.

Visualization of large and complex networks is needed in many applications such as biology, social science, and software engineering. A good visualization reveals the hidden structure of a network, highlights patterns and trends, makes it easy to see outliers. Thus a good visualization leads to new insights, findings and predictions.
However, visualizing large and complex real-world networks is challenging, because most existing graph drawing algorithms mainly focus on planar graphs, and consequently have made little impact on visualization of real-world complex networks, which are non-planar.
Therefore, effective drawing algorithms for beyond-planar graphs are in high demand from industry and other application domains.

The notion of  $1$-planarity  is among the most natural and most studied generalizations of planarity.   A graph is $1$-planar if there exists an embedding in which every edge is crossed at most once.
The family of $1$-planar graphs was introduced by Ringel~\cite{R65} in 1965 in the context of the simultaneous vertex-face coloring of planar graphs. Specifically, if we consider a given planar graph and its dual graph and add edges between the primal vertices and the dual vertices, we obtain a $1$-planar graph. Ringel was interested in a generalization of the 4-color theorem for planar graphs, he proved that every $1$-planar graph has chromatic number at most 7, and conjectured that this bound could be lowered to 6~\cite{R65}. Borodin~\cite{B84} settled the conjecture in the affirmative, proving that the chromatic number of each $1$-planar graph is at most $6$ (the bound is tight as for example the complete graph $K_6$ is $1$-planar and requires six colors).

The aim of this paper is to present an annotated bibliography of papers devoted to the study of combinatorial properties,  geometric properties, and algorithms for $1$-planar graphs. Similar to~\cite{BattistaETT94}, this annotated bibliography reports the references in the main body of the sections; we believe that this choice can better guide the reader through the different results, while reading the different sections. All references are also collected at the end of the paper, in order to give the reader an easy access to a complete bibliography.

\paragraph{Paper organization}
The remainder of this paper is structured as follows.
\begin{itemize}

\item Section~\ref{se:preliminaries} contains basic terminology, notation, and definitions. It is divided in two main parts as follows. Section~\ref{sse:basic} introduces basic definitions and notation about graphs, drawings, and embeddings; this section can be skipped by  readers familiar with graph theoretic concepts and graph drawing. Section~\ref{sse:1planar} contains more specific definitions for $1$-planar graphs and for subclasses of $1$-planar graphs such as IC-planar and NIC-planar graphs. Because different papers on $1$-planarity often use different terminology and notation, this section is important for the rest of the paper. 

\item Section~\ref{se:characterization-rec} is concerned with two fundamental and closely related aspects of $1$-planar graphs. Specifically, Section~\ref{sse:characterization} and Section~\ref{sse:recognition} survey known results for the problem of characterizing and recognizing $1$-planar graphs, respectively. While the recognition problem is NP-complete for general $1$-planar graphs, interesting characterizations and polynomial-time recognition algorithms are known for some meaningful classes of $1$-planar graphs, such as optimal $1$-planar graphs.

\item Section~\ref{se:properties} investigates structural properties of $1$-planar graphs, i.e., properties that depend only on the abstract structure of these graphs. Specifically, this section contains: the main results on the chromatic number, the chromatic index, and on other coloring parameters (Section~\ref{sse:coloring}); upper and lower bounds on the edge density for several classes of $1$-planar graphs (Section~\ref{sse:density}); results on edge partitions with specific properties on the edge sets (Section~\ref{sse:edgepartitions}); bounds on various graph parameters (Section~\ref{sse:parameters}); results on the existence of subgraphs with bounded vertex degree (Section~\ref{sse:subgraphs}); results on binary graph operations that preserve $1$-planarity (Section~\ref{sse:operations}).

\item Section~\ref{se:geometric-representations} describes several results about geometric representations of $1$-planar graphs, which are of interest in computational geometry, graph drawing and network visualization. In particular, the following types of representation are considered: straight-line drawings (Section~\ref{sse:straightline});  drawings with right-angle crossings (Section~\ref{sse:rac}); visibility representations (Section~\ref{sse:visibility}); and contact representations (Section~\ref{sse:contact}).

\end{itemize}

\section{Preliminaries}\label{se:preliminaries}

\subsection{Basic definitions}\label{sse:basic}

In this section we introduce some preliminaries and definitions about graphs and about drawings and embeddings of graphs. For more details about graph theory, the reader is referred to classic textbooks such as:

\begin{itemize}
\item \cite{bm-gt-07}~J.~A.~Bondy and U.~S.~R.~Murty.
\newblock {\em Graph theory}.
\newblock Graduate texts in mathematics.~Springer, 2007.
\item \cite{Diestel12}~R.~Diestel.
\newblock {\em Graph Theory, 4th Edition}, volume 173 of {\em Graduate texts in
  mathematics}.
\newblock Springer, 2012.
\item \cite{g-agt-85}~A.~Gibbons.
\newblock {\em {Algorithmic Graph Theory}}.
\newblock Cambridge University Press, 1985.
\item \cite{h-gt-72}~F.~Harary, editor.
\newblock {\em Graph Theory}.
\newblock Addison-Wesley, 1972.
\end{itemize}

Further graph drawing background can also be obtained in several books and surveys on the topic, including:
\begin{itemize}
\item \cite{dett-gdavg-99}~G.~{Di Battista}, P.~Eades, R.~Tamassia, and I.~G.~Tollis.
\newblock {\em Graph Drawing: Algorithms for the Visualization of Graphs}.
\newblock Prentice-Hall, 1999.

\item \cite{Juenger04}~M.~J{\"{u}}nger and P.~Mutzel, editors.
\newblock {\em Graph Drawing Software}.
\newblock Springer, 2004.

\item \cite{kw-dgmm-01}~M.~Kaufmann and D.~Wagner, editors.
\newblock {\em Drawing Graphs, Methods and Models}, volume 2025 of {\em LNCS}. Springer, 2001.

\item \cite{NR04}~T.~Nishizeki and M.~S.~Rahman.
\newblock {\em Planar graph drawing}, volume~12 of {\em Lecture notes series on   computing}. 
\newblock World Scientific, 2004.

\item \cite{Sugiyama02}~K.~Sugiyama.
\newblock {\em Graph Drawing and Applications for Software and Knowledge  Engineers}, volume~11 of {\em Series on Software Engineering and Knowledge  Engineering}.
\newblock World Scientific, 2002.

\item \cite{Tamassia99}~R.~Tamassia.
\newblock Advances in the theory and practice of graph drawing.
\newblock {\em Theor.~Comput.~Sci.}, 217(2):235--254, 1999.

\item \cite{2013gd}~R.~Tamassia, editor.
\newblock {\em Handbook on Graph Drawing and Visualization}.
\newblock Chapman and Hall/CRC, 2013.
\end{itemize}

In order to make this paper self-contained, we review several basic definitions that will be used in the next sections.

\subsubsection{Graphs}

A \emph{graph} is a mathematical discrete structure used to represent a set of interconnected objects.  Formally, a graph $G=(V,E)$ is an ordered pair comprising a finite and non-empty set $V$ of elements called \emph{vertices}  and a finite set $E$ of elements called \emph{edges}. An edge $e \in E$ is an unordered pair $(u,v)$ of vertices; vertices $u$ and $v$ are called \emph{end-vertices} of $e$. We often write $e=(u,v)$ to denote an edge $e$ with end-vertices $u$ and $v$. Vertices $u$ and $v$ are said to be \emph{adjacent}, while $e$ is \emph{incident} to all of its end-vertices. The set of vertices adjacent to a vertex $v$ is called the \emph{neighborhood} of $v$ and denoted by $N(v)$. The number of edges that are incident to a vertex $v$ is called the \emph{degree} of $v$ and is denoted by $deg(v)$. A graph $G$ has maximum vertex degree $\Delta$, if $deg(v) \le \Delta$ for every vertex $v$ of $G$ and there is a vertex $u$ of $G$ such that $deg(u) = \Delta$. An edge of  $G$ is a \emph{self-loop} if its end-vertices coincide. Also, $G$ contains \emph{multiple edges} if it has two or more edges with the same end-vertices. A graph $G$ is \emph{simple} if it has neither self-loops nor multiple edges. Note that in a simple graph the degree of a vertex $v$ corresponds to the size of $N(v)$.  In the following we always refer to simple graphs, unless specified otherwise.

A subgraph $G'$ of a graph $G=(V,E)$ is a graph $G'=(V',E')$, such that $V' \subseteq V$ and $E' \subseteq E$. If $V'$ is a subset of $V$, the subgraph of $G$ \emph{induced} by $V'$ is the graph $G'=(V', E')$, where $E' \subseteq E$ is the subset of all edges in $E$ connecting any two vertices that are in $V'$. A subgraph $G'=(V',E')$ of $G=(V,E)$ such that $V'=V$ is a \emph{spanning subgraph} of $G$.

A \emph{path} is a graph whose set of vertices is $V = \{v_1,\dots,v_n\}$ and there exists an edge $e_i = (v_i,v_{i+1})$ for $i=1,\dots,n-1$.

A \emph{component} $G'$ of a graph $G$ is a maximal subgraph of $G$ such that for every pair $u$, $v$ of vertices of $G'$ there is a path between $u$ and $v$ in $G'$. A graph that has exactly one component is said to be \emph{connected}.  A \emph{separating $k$-set}, $k\geq1$, of a graph $G$ is a set of $k$ vertices whose removal increases the number of connected components of $G$. Separating $1$-sets and $2$-sets are called \emph{cut vertices} and \emph{separation pairs}, respectively. A connected graph is  \emph{$2$-connected} if it has no cut vertices. A $2$-connected graph is  \emph{$3$-connected} if it has no separation pairs. In general, a graph is \emph{$k$-connected}, $k\geq2$, if it has no separating ($k-1$)-sets. Clearly, if a graph is \emph{$k$-connected}, then it is also ($k-1$)-\emph{connected}.

A connected graph $G$ is a \emph{cycle} if every vertex of $G$ has degree $2$. A graph is \emph{acyclic} if it does not contain subgraphs that are cycles. A connected acyclic graph is called a \emph{tree}.
A collection of trees (i.e., an acyclic graph) is a \emph{forest}.

A graph is \emph{complete} if it has an edge connecting every pair of vertices. A complete graph with $n$ vertices is denoted by $K_n$. A graph $G$ is \emph{$k$-partite} if the set of its vertices can be partitioned into $k \ge 2$ sets (also called \emph{parts}), $V_1,\dots,V_k$, such that every edge of $G$ connects a vertex in $V_i$ to a vertex in $V_j$, for some $1 \le i \ne j \leq k$. A graph $G$ is \emph{complete $k$-partite} if it is $k$-partite and every two vertices that belong to distinct parts are adjacent. A 2-partite graph is also called a \emph{bipartite graph}. A complete bipartite graph such that $|V_1|=n_1$ and $|V_2|=n_2$ is denoted by $K_{n_1,n_2}$.

\subsubsection{Drawings and Embeddings}

Let $G=(V,E)$ be a graph,  a \emph{drawing on the plane}, or simply a \emph{drawing}, $\Gamma$ of $G$ is a pair of functions $\langle \Gamma_V,\Gamma_E \rangle$, where $\Gamma_V$ maps each vertex $v$ to a point $\Gamma_V(v)$ of the plane and $\Gamma_E$ maps each edge $(u,v)$ to a simple open Jordan curve $\Gamma_E(u,v)$, whose endpoints are $\Gamma_V(u)$ and $\Gamma_V(v)$. The curves representing the edges are allowed to intersect, but they may not pass through vertices except for their endpoints. In what follows, we will often refer to the vertices and to the edges of a drawing as an abbreviation for the points and the curves that represent the vertices and the edges of the depicted graph.

A drawing is \emph{planar} if there are no two edges that cross (except at common endpoints). A \emph{planar graph} is a graph that admits a planar drawing.

A drawing divides the plane into topologically connected regions, called \emph{faces}. The infinite region is called the \emph{outer face}. For a planar drawing the boundary of a face consists of vertices and edges, while for a non-planar drawing the boundary of a face may contain vertices, crossings, and edges (or parts of edges). Given a drawing $\Gamma$, the \emph{planarization} $\Gamma_p$ of $\Gamma$ is the drawing obtained by replacing each crossing point of $\Gamma$ with a \emph{dummy vertex}. The vertices of $\Gamma$ in $\Gamma_p$ are called \emph{original vertices} to avoid confusion.

A drawing where all the edges are mapped to segments is called a \emph{straight-line drawing}. A \emph{polyline drawing} is a drawing where the edges are mapped to chains of segments. A drawing, either straight-line or polyline, where all the vertices and all the possible bend points are mapped to points with integer coordinates is a \emph{grid drawing}.
The \emph{bounding box} of a grid drawing $\Gamma$ is the minimum axis-aligned box containing $\Gamma$. If the bounding box has side lengths $X-1$ and $Y-1$, then we say that $\Gamma$ has \textit{area} $X \times Y$.

A \emph{rotation system} of a graph is defined by an assignment of a clockwise order of the edges incident to a vertex, for all vertices of the graph.  A drawing is said to \emph{respect} a rotation system if scanning around a vertex in clockwise order encounters the edges in the prescribed order.

An \emph{embedding} of a graph $G$ is an equivalence class of drawings of $G$ that define the same set of faces and the same outer face. A \emph{planar embedding} is an embedding that represents an equivalence class of planar drawings. Equivalently, an embedding of a graph is an equivalence class of drawings under homeomorphisms of the plane. 
For planar graphs the information carried in an embedding is equivalent to that in a rotation system together with one face indicated to be the outer face.
More in general, a graph with a \emph{fixed embedding} comes with a rotation system, a clockwise order of (part of) edges around each crossing, and with one face indicated to be the outer face. A \emph{plane graph} is a graph with a fixed planar embedding. Similarly as for a drawing, given a graph $G$ with a fixed embedding, the \emph{planarization} $G_p$ of $G$ is the graph obtained by replacing each crossing of $G$ with a \emph{dummy vertex}. The vertices of $G$ in $G_p$ are called \emph{original vertices} to avoid confusion.   The planarization $G_p$ is a \emph{plane} graph, i.e., a planar graph with a fixed embedding without crossings. It is worth recalling that a $3$-connected planar graph has a unique planar embedding up to the choice of the outer face~\cite{whitney32}.

\begin{itemize}
\item \cite{whitney32}~H.~Whitney.
\newblock Congruent graphs and the connectivity of graphs.
\newblock {\em Am. J. Math.}, 54(1):150--168, 1932.
\end{itemize}

\subsection{1-planar Graphs}\label{sse:1planar}

A \emph{$1$-planar drawing} is one in which each edge is crossed at most once. A graph is \emph{$1$-planar} if it admits a $1$-planar drawing. A \emph{$1$-planar embedding} is an embedding that represents an equivalence class of $1$-planar drawings. A \emph{$1$-plane graph} is a graph with a fixed $1$-planar embedding.

A $1$-planar graph $G$ with $n$ vertices has at most $4n-8$ edges~\cite{BSW83,PachT97} (see also Section~\ref{sse:density}). If $G$ has exactly $4n-8$ edges, then it is called \emph{optimal}. A $1$-planar graph $G$ is \emph{maximum} if it has the maximum number of edges over all $1$-planar graphs with $n$ vertices. Note that there exist graphs that are maximum but not optimal (e.g., the complete graph with 5 vertices), whereas an optimal $1$-planar graph is also maximum. A $1$-planar graph $G$ is \emph{maximal} if no edge can be added to $G$ without having at least one edge crossed twice in every drawing of $G$. A $1$-planar graph is \emph{triangulated} if it admits a drawing in which every face contains either exactly three distinct vertices or exactly two distinct vertices and one crossing point.

\begin{itemize}

\item \cite{BSW83}~R.~Bodendiek, H.~Schumacher, and K.~Wagner.
\newblock Bemerkungen zu einem {S}echsfarbenproblem von G.~Ringel.
\newblock {\em Abhandlungen aus dem Mathematischen Seminar der Universitaet  Hamburg}, 53(1):41--52, 1983.

\item \cite{PachT97}~J.~Pach and G.~T{\'o}th.
\newblock Graphs drawn with few crossings per edge.
\newblock {\em Combinatorica}, 17(3):427--439, 1997.

\end{itemize}

Let $\Gamma$ be a $1$-planar drawing, and let $\Gamma_p$ be its planarization. The drawing $\Gamma$ is of of class $C_x$, for $x \in \{0,1,2\}$, if for every two dummy vertices $u_1$ and $u_2$, it holds that $|N(u_1) \cap N(u_2)| \le x$. A $1$-planar graph $G$ is of class $C_x$, for $x \in \{0,1,2\}$, if it has a drawing of class $C_x$ and no drawing of class $C_{x-1}$ (if $x>0$). A $1$-planar graph of class $C_0$ is also called an \emph{IC-planar graph}, where IC stands for \emph{independent crossings} (see, e.g.,~\cite{AMC10}). A $1$-planar graph of class $C_1$ is also called a \emph{NIC-planar graph}, where NIC stands for \emph{near-independent crossings} (see, e.g.,~\cite{Zhang2014}). Czap  and {\v S}ugerek~\cite{CzapS17} observed that every $1$-planar graph belongs to one of the classes $C_0$, $C_1$, or $C_2$. The concepts of \emph{IC-planar} and \emph{NIC-planar embedding}, and of \emph{IC-plane} and \emph{NIC-plane graph}, can be defined similarly as for the $1$-planar case.

\begin{itemize}

\item \cite{AMC10} M.~O.~Albertson.
\newblock Chromatic number, independence ratio, and crossing number.
\newblock {\em Ars Math.~Contemp.}, 1(1), 2008.

\item \cite{CzapS17} J.~{Czap} and P.~{{\v S}ugerek}.
\newblock Drawing graph joins in the plane with restrictions on crossings.
\newblock {\em Filomat}, 31(2):363–--370, 2017.

\item \cite{Zhang2014} X.~Zhang.
\newblock Drawing complete multipartite graphs on the plane with restrictions  on crossings.
\newblock {\em Acta Math.~Sin.~English Series}, 30(12):2045--2053, 2014.

\end{itemize}

A drawing is \emph{outer $1$-planar} if it is $1$-planar and all the vertices belong to the outer face of the drawing. An \emph{outer $1$-planar graph} is a graph that admits an outer $1$-planar drawing. One can define the concepts of \emph{outer $1$-planar embedding} and \emph{outer $1$-plane graph} similarly as above.

\section{Characterization and Recognition}\label{se:characterization-rec}

In this section we deal with the problems of characterizing and recognizing $1$-planar graphs. We recall that finding a \emph{characterization} for a family $\mathcal F$ of graphs means finding a property $P$ such that a graph $G$ has the property $P$ if and only if $G \in \mathcal F$, while the \emph{recognition problem} for a family of graphs $\mathcal F$ is the problem of deciding whether a graph $G$ belongs to $\mathcal F$. The two problems are closely related. In particular, if a property $P$ can be verified in polynomial time on a graph $G$ and $P$ characterizes a family of graphs $\mathcal F$, then deciding whether $G \in \mathcal F$ can be done in polynomial time. We investigate these two problems for general $1$-planar graphs and for several subfamilies of $1$-planar graphs.

\subsection{Characterization}\label{sse:characterization}

In what follows we survey results concerning the characterization of general $1$-planar graphs (Section~\ref{ssse:general}), optimal $1$-planar graphs (Section~\ref{ssse:optimal}), triangulated $1$-planar graphs (Section~\ref{ssse:triangulated}), and complete $k$-partite $1$-planar graphs (Section~\ref{ssse:complete-partite}).

\subsubsection{General 1-planar graphs}\label{ssse:general}
 The class of $1$-planar graphs is not closed under edge-contraction, and thus $1$-planar graphs are not minor closed~\cite{ChenK05}. For example, the  $n \times n \times 2$ grid graph which is $1$-planar and contains $K_n$ as a minor, see~\cite{2015arXiv150604380D}.  

\begin{itemize}

\item \cite{ChenK05}~Z.~Chen and M.~Kouno.
\newblock A linear-time algorithm for 7-coloring $1$-plane graphs.
\newblock {\em Algorithmica}, 43(3):147--177, 2005.

\item \cite{2015arXiv150604380D}~V.~{Dujmovi{\'c}}, D.~{Eppstein}, and D.~R. {Wood}.
\newblock {Structure of Graphs with Locally Restricted Crossings}.
\newblock {\em CoRR}, abs/1506.04380, 2015.
\newblock To appear in J. Discr. Math.

\end{itemize}

A \emph{minimal non-$1$-planar graph} is a  graph that is not $1$-planar and such that all its proper subgraphs are $1$-planar. Korzhik and Mohar~\cite{KM13} proved that the number of  $n$-vertex distinct (i.e., non-isomorphic) minimal non-$1$-planar graphs is exponential in $n$ (for $n \ge 63$).  These results can be seen as an indication that a characterization of $1$-planar graphs by Wagner-type theorem using a finite number of forbidden  minors is unlikely. Similarly, for any graph there exists a $1$-planar subdivision, which implies the impossibility of having a  characterization of $1$-planar graphs by Kuratowski-type theorem using a finite number of forbidden  subdivisions.

\begin{itemize}

\item \cite{KM13}~V.~P.~Korzhik and B.~Mohar.
\newblock Minimal obstructions for 1-immersions and hardness of $1$-planarity  testing.
\newblock {\em J.~Graph Theory}, 72(1):30--71, 2013.

\end{itemize}

\subsubsection{Optimal 1-planar graphs}\label{ssse:optimal} 
We recall that an $n$-vertex $1$-planar graph is optimal if it contains $4n-8$ edges (see also Section~\ref{sse:1planar}).
\begin{figure}
\centering
\subfigure[$Q$]{\includegraphics[scale=0.8,page=1]{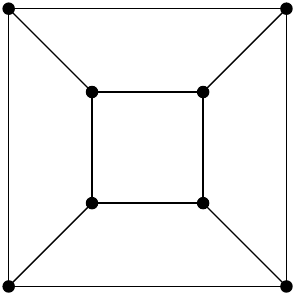}\label{fi:quadrangulation}}\hfil
\subfigure[$G$]{\includegraphics[scale=0.8,page=2]{figs/optimal1planar}\label{fi:optimal1planar}}
\caption{(a) A $3$-connected planar quadrangulation $Q$ and (b) the optimal $1$-planar graph $G$ constructed by inserting a pair of crossing edges in each face of $Q$.\label{fi:o1p}}
\end{figure}
A \emph{planar quadrangulation} is a planar graph that admits a drawing in which each face $f$ has four vertices. Let $Q$ be a $3$-connected planar quadrangulation. Since $Q$ is $3$-connected, it has a unique planar embedding up to the choice of the outer face, and thus a uniquely defined set of faces (or, equivalently, a unique planar embedding on the sphere). Let $f$ be a face of $Q$, such that in a closed walk along its boundary we encounter its four vertices in this order: $u_1,u_2,u_3,u_4$. The operation of \emph{inserting a pair of crossing edges in $f$} means adding the edges $(u_1,u_3)$ and $(u_2,u_4)$ inside $f$ such that they cross each other. Note that the two edges do not exist in $G$ (since $G$ is bipartite). Note also that the resulting graph is clearly $1$-planar.  Bodendiek \emph{et al.}~\cite{BSW83,BSW84} (and later Suzuki~\cite{S10}) proved that a $1$-planar graph $G$ is optimal if and only if  $G$ can be obtained from a $3$-connected planar quadrangulation $Q$ by inserting a pair of crossing edges inside each face of $Q$; see Figure~\ref{fi:o1p} for an illustration. Based on this characterization, it is possible to prove that there are no $n$-vertex optimal $1$-planar graphs with $n \le 7$ or $n=9$ vertices, while $n$-vertex optimal $1$-planar graphs always exist with $n=8$ and $n \ge 10$ vertices~\cite{BSW83,BSW84}.

\begin{itemize}
\item \cite{BSW83}~R.~Bodendiek, H.~Schumacher, and K.~Wagner.
\newblock Bemerkungen zu einem {S}echsfarbenproblem von G.~Ringel.
\newblock {\em Abhandlungen aus dem Mathematischen Seminar der Universitaet  Hamburg}, 53(1):41--52, 1983.

\item \cite{BSW84}~R.~Bodendiek, H.~Schumacher, and K.~Wagner.
\newblock Uber 1-optimale Graphen.
\newblock {\em Mathematische Nachrichten}, 117(1):323--339, 1984.

\item \cite{S10}~Y.~Suzuki.
\newblock Optimal $1$-planar graphs which triangulate other surfaces.
\newblock {\em Discrete Math.}, 310(1):6--11, 2010.
\end{itemize}

\begin{figure}[t]
\centering
\subfigure[]{\includegraphics[scale=0.75,page=3]{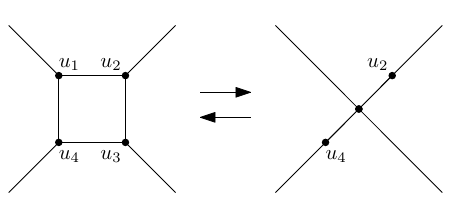}\label{fi:o1p-operations-1}}\hfil
\subfigure[]{\includegraphics[scale=0.75,page=4]{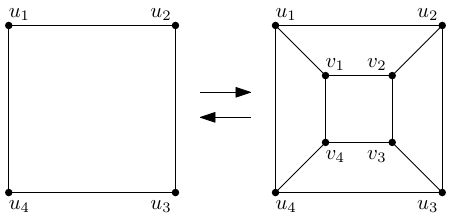}\label{fi:o1p-operations-2}}
\\
\subfigure[]{\includegraphics[scale=0.75,page=5]{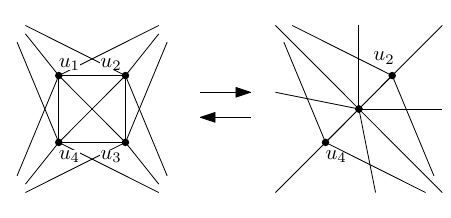}\label{fi:o1p-operations-3}}\hfil
\subfigure[]{\includegraphics[scale=0.75,page=6]{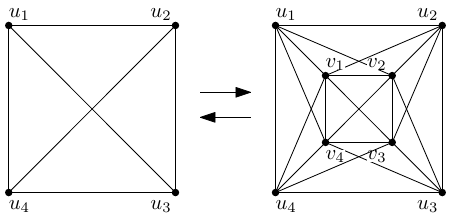}\label{fi:o1p-operations-4}}
\\
\subfigure[$W_{8}$]{\includegraphics[scale=0.75,page=7]{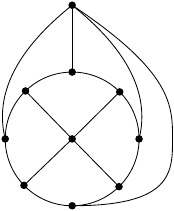}\label{fi:o1p-pdoublewheel}}\hfil
\subfigure[$XW_{8}$]{\includegraphics[scale=0.75,page=8]{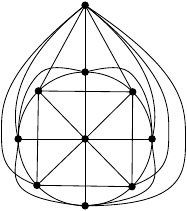}\label{fi:o1p-xpdoublewheel}}
\caption{The face contraction and vertex splitting operations for (a) planar quadrangulations and (c) optimal $1$-planar graphs. The 4-cycle addition and 4-cycle removal operations for (b) planar quadrangulations and (d) optimal $1$-planar graphs. (e) The pseudo double wheel $W_{8}$. (f) The $X$-pseudo double wheel $XW_{8}$. \label{fi:o1p-charcterization}}
\end{figure}

The above characterization establishes a one-to-one mapping between the set of $3$-connected planar quadrangulations and the set of optimal $1$-planar graphs. A similar family of $1$-planar graphs are the \emph{kinggraphs}~\cite{Chepoi2002}. They are defined as those $1$-planar graphs that can be generated by inserting a pair of crossing edges in each quadrangular inner face of a square-graph (i.e., a plane graph in which all inner faces have four vertices and every such vertex  has degree at least $4$).  These $1$-planar graphs generalize the   graphs that represent all legal moves of the king chess piece on a chessboard, called \emph{King's graphs}~\cite{Chang2013}.

Let $Q$ be a $3$-connected planar quadrangulation. Let $f$ be a face of $Q$ such that in a closed walk along its boundary we encounter its four vertices in this order: $u_1,u_2,u_3,u_4$. The operation of \emph{face contraction} of $f$ at $\{u_1,u_3\}$ in $Q$ identifies $u_1$ and $u_3$ and replaces the two pairs of multiple edges $(u_1,u_2),(u_2,u_3)$ and $(u_1,u_4),(u_3,u_4)$ with two single edges, respectively. The inverse operation is called a \emph{vertex-splitting}; see Figure~\ref{fi:o1p-operations-1} for an illustration. These two operations can be applied only if they do  not break the simplicity or the triconnectivity of the graph. A \emph{4-cycle addition}  to $f$ is the operation of adding a 4-cycle $v_1,v_2,v_3,v_4$ inside $f$ and adding the crossing-free edges $(u_i,v_i)$, for $i=1,2,3,4$. The inverse operation is called \emph{4-cycle removal}; see Figure~\ref{fi:o1p-operations-2} for an illustration. This last operation can be applied only if it  does not break the triconnectivity of the graph. A planar quadrangulation is \emph{irreducible} if neither face contraction nor 4-cycle removal can be applied to $H$.

The \emph{pseudo double wheel} $W_{2k}$ is a $3$-connected planar quadrangulation with $2k+2$ vertices, while the \emph{$X$-pseudo double wheel} $XW_{2k}$  is the corresponding optimal $1$-planar graph. See Figure~\ref{fi:o1p-pdoublewheel} for an illustration of the pseudo double wheel when $k=4$, and Figure~\ref{fi:o1p-xpdoublewheel} for an illustration of the $X$-pseudo double wheel when $k=4$. Brinkmann \emph{et al.}~\cite{BGGMTW05} proved that $W_{2k}$ ($k \ge 3$) is the unique series of irreducible $3$-connected planar quadrangulations, and that every $3$-connected planar quadrangulation can be obtained from the graph $W_{2k}$ ($k \ge 3$) by a sequence of vertex-splitting and 4-cycle additions. The four operations defined above for $3$-connected planar quadrangulations can be easily extended to optimal $1$-planar graphs. See also Figures~\ref{fi:o1p-operations-3} and~\ref{fi:o1p-operations-4} for an illustration. In particular, Suzuki~\cite{S10b} observed that every optimal $1$-planar graph can be obtained from the graph $W_{2k}$ ($k \ge 3$) by a sequence of vertex-splitting and 4-cycle additions. This characterization has been used by Suzuki~\cite{S10} to investigate the re-embeddability of optimal $1$-planar graphs.

\begin{itemize}

\item \cite{BGGMTW05}~G.~Brinkmann, S.~Greenberg, C.~S.~Greenhill, B.~D.~McKay, R.~Thomas, and P.~Wollan.
\newblock Generation of simple quadrangulations of the sphere.
\newblock {\em Discrete Math.}, 305(1--3):33--54, 2005.

\item \cite{Chang2013}~G.~J. Chang.
\newblock {\em Algorithmic Aspects of Domination in Graphs}, pages 221--282.\newblock Springer, New York, NY, 2013.

\item \cite{Chepoi2002}~V.~Chepoi, F.~Dragan, and Y.~Vax\`{e}s.
\newblock Center and diameter problems in plane triangulations and  quadrangulations. \newblock In {\em {ACM-SIAM SODA 2002}}, pages 346--355. SIAM, 2002.

\item \cite{S10b}~Y.~Suzuki.
\newblock Optimal $1$-planar graphs which triangulate other surfaces.
\newblock {\em Discrete Math.}, 310(1):6--11, 2010.

\item \cite{S10}~Y.~Suzuki.
\newblock Re-embeddings of maximum $1$-planar graphs.
\newblock {\em {SIAM} J.~Discrete Math.}, 24(4):1527--1540, 2010.

\end{itemize}

\subsubsection{Triangulated 1-planar graphs}\label{ssse:triangulated}
A \emph{map} $M$ is a partition of the sphere into finitely many regions. Each region is a closed disk and the interiors of two regions are disjoint. Some regions are labeled as \emph{countries} and the non-labeled regions are called \emph{holes}. Two countries are \emph{adjacent} if they touch. Given a graph $G$, a map $M$ \emph{realizes} $G$ if the vertices of $G$ are in bijection with the countries of $M$, and two countries of $M$ are adjacent if and only there is an edge between the two corresponding vertices of $G$. A graph that can be realized as a map is called a \emph{map graph}. Note that $h>1$ countries meeting at a point implies the existence of a complete graph $K_h$ as a subgraph of $G$. If no more than $h$ regions meet at a point, then $M$ is an \emph{$h$-map} and $G$ an \emph{$h$-map graph}. Also, if a map $M$ has no holes, then it is called a \emph{hole-free map} and its map graph $G$ is a \emph{hole-free map graph}. We refer the reader to papers by Chen \emph{et al.}~\cite{ChenGP98,ChenGP02,ChenGP06} and by Thorup~\cite{Thorup98} for more details and results on this family of graphs. In particular, Chen \emph{et al.}~\cite{ChenGP06} observed that $3$-connected hole-free 4-map graphs are exactly the triangulated $1$-planar graphs, that is, a $3$-connected graph is triangulated $1$-planar if and only if it can be realized as a hole-free 4-map. Note that the family of $1$-planar graphs includes the family of 4-map graphs, but the opposite is not true~\cite{ChenGP02}. The relationship between $1$-planar graphs and map graphs has been further investigated by Brandenburg~\cite{Brandenburg15}.

\begin{itemize}

\item \cite{Brandenburg15}~F.~J.~Brandenburg.
\newblock On 4-map graphs and $1$-planar graphs and their recognition problem.
\newblock {\em CoRR}, abs/1509.03447, 2015.

\item \cite{ChenGP98}~Z.~Chen, M.~Grigni, and C.~H.~Papadimitriou.
\newblock Planar map graphs.
\newblock In {\em {STOC} 1998}, pages 514--523. {ACM}, 1998.

\item \cite{ChenGP02}~Z.~Chen, M.~Grigni, and C.~H.~Papadimitriou.
\newblock Map graphs.
\newblock {\em J.~{ACM}}, 49(2):127--138, 2002.

\item \cite{ChenGP06}~Z.~Chen, M.~Grigni, and C.~H.~Papadimitriou.
\newblock Recognizing hole-free 4-map graphs in cubic time.
\newblock {\em Algorithmica}, 45(2):227--262, 2006.

\item \cite{Thorup98}~M.~Thorup.
\newblock Map graphs in polynomial time.
\newblock In {\em {FOCS} 1998}, pages 396--405. {IEEE}, 1998.

\end{itemize}

\subsubsection{Complete $k$-partite 1-planar graphs}\label{ssse:complete-partite}
While it is known that the complete graph $K_n$ is $1$-planar if and only if $n \le 6$, Czap and Hud{\'a}k~\cite{CzapH12} characterized the complete
$k$-partite $1$-planar graphs. 
Specifically, $K_{n_1,n_2}$ is not $1$-planar for  $n_1,n_2 \ge 5$,  $K_{n_1,3}$ is $1$-planar if and only if $n_1 \le 6$, and $K_{n_1,4}$ is $1$-planar if and only if $n_1 \le 4$ (recall that $K_{n_1,1}$ and $K_{n_1,2}$ are planar for all values of $n_1$). For $h \ge 7$, the graph $K_{n_1,\dots,n_h}$ contains $K_7$ as a subgraph, hence it cannot be $1$-planar. For $3 \le h \le 6$, the characterization is based on case analysis and we let the reader refer to~\cite{CzapH12} for details. 

\begin{itemize}

\item \cite{CzapH12}~J.~Czap and D.~Hud{\'{a}}k.
\newblock $1$-planarity of complete multipartite graphs.
\newblock {\em Discrete Appl.~Math.}, 160(4-5):505--512, 2012.

\end{itemize}

\subsection{Recognition}\label{sse:recognition}

\newcommand{\oneplanarity}{$1$-planarity problem\xspace}
In what follows we refer to the problem of deciding whether a graph $G$ is $1$-planar as the \emph{\oneplanarity}. In general this problem in NP-hard (Section~\ref{ssse:hardness}), although it can be solved in polynomial time in some restricted cases (Section~\ref{ssse:polytime}). \oneplanarity  has been studied also in the setting where the input graph has a fixed rotation system (Section~\ref{ssse:fixed-rotation}), and in terms of parameterized complexity (Section~\ref{ssse:parameterized}). 

\subsubsection{NP-hardness results}\label{ssse:hardness}
The \oneplanarity is NP-complete~\cite{GB07,KM13}. The first proof was by Grigoriev and Boadlander~\cite{GB07}, and was based on a  reduction from the \emph{3-partition problem}. We recall that the 3-partition problem is an NP-complete problem defined	as follows (see also~\cite{GareyJ79}). Given a set $A$ of $3m$ elements, an integer $B\in\mathbb N$, and an integer $s(a)\in\mathbb N$ for each $a\in A$ such that $B/4<s(a)<B/2$ and $\sum_{a\in A}s(a)=mB$, the problem asks if $A$ can be partitioned into $m$ subsets $A_1,...,A_m$ such that, for $1\leq i\leq m$, $\sum_{a\in A_i}s(a)=B$. On a high level, the reduction works by constructing a graph $G$ that separates the plane into $m$ large regions $R_1, \dots, R_m$, such that the boundaries between each
region cannot be crossed in a $1$-planar drawing of $G$. Each region $R_i$ has 3 buckets to place the elements of $A_i$, and $B$ buckets to place the elements of $A$.

A second proof was later given by Korzhik and Mohar~\cite{KM13}, who used a reduction from the \emph{3-coloring problem for planar graphs} (see~\cite{GareyJ79})). 

\begin{itemize}

\item \cite{GareyJ79}~M.~R.~Garey and D.~S.~Johnson.
\newblock {\em Computers and Intractability: {A} Guide to the Theory of  {NP}-Completeness}.
\newblock W.~H.~Freeman, 1979.

\item \cite{GB07}~A.~Grigoriev and H.~L.~Bodlaender.
\newblock Algorithms for graphs embeddable with few crossings per edge.
\newblock {\em Algorithmica}, 49(1):1--11, 2007.

\item \cite{KM13}~V.~P.~Korzhik and B.~Mohar.
\newblock Minimal obstructions for 1-immersions and hardness of $1$-planarity  testing.
\newblock {\em J.~Graph Theory}, 72(1):30--71, 2013.

\end{itemize}

The \oneplanarity remains NP-complete even for graphs with bounded bandwidth, pathwidth, or treewidth~\cite{BCE13}, and for graphs obtained from planar graphs by adding a single edge~\cite{CM12}.

\begin{itemize}

\item \cite{BCE13}~M.~J.~Bannister, S.~Cabello, and D.~Eppstein.
\newblock Parameterized complexity of $1$-planarity.
\newblock In {\em {WADS} 2013}, volume 8037 of {\em LNCS}, pages 97--108.  Springer, 2013.

\item \cite{CM12}~S.~Cabello and B.~Mohar.
\newblock Adding one edge to planar graphs makes crossing number and  $1$-planarity hard.
\newblock {\em {SIAM} J.~Comput.}, 42(5):1803--1829, 2013.

\end{itemize}

Brandenburg \emph{et al.}~\cite{BrandenburgDEKL16} proved that the problem of deciding whether a graph $G$ is IC-planar is also NP-complete, by using a reduction from the \oneplanarity. A similar reduction holds for NIC-planar graphs as well~\cite{BachmaierBHNR17}.

\begin{itemize}

\item \cite{BachmaierBHNR17}~C.~Bachmaier, F.~J.~Brandenburg, K.~Hanauer, D.~Neuwirth, and J.~Reislhuber.
\newblock NIC-planar graphs.
\newblock {\em CoRR}, abs/1701.04375, 2017.

\item \cite{BrandenburgDEKL16}~F.~J.~Brandenburg, W.~Didimo, W.~S.~Evans, P.~Kindermann, G.~Liotta, and
  F.~Montecchiani.
\newblock Recognizing and drawing {IC}-planar graphs.
\newblock {\em Theor.~Comput.~Sci.}, 636:1--16, 2016.

\end{itemize}

\subsubsection{Polynomial-time results}\label{ssse:polytime}
Brandenburg~\cite{Brandenburg16a} recently showed that the characterization for optimal $1$-planar graphs given by Suzuki~\cite{S10b} (see Section~\ref{sse:characterization}) can be exploited to efficiently recognize optimal $1$-planar graphs. Specifically, given an $n$-vertex graph with $4n-8$ edges, there exists an $O(n)$-time algorithm to decide whether $G$ is optimal $1$-planar and if $G$ is optimal $1$-planar, then the algorithm returns a valid embedding of $G$.

\begin{itemize}

\item \cite{Brandenburg16a}~F.~J.~Brandenburg.
\newblock Recognizing optimal 1-planar graphs in linear time.
\newblock {\em Algorithmica}, 2016.

\end{itemize}

Triangulated $1$-planar graphs can be recognized in polynomial time, since these graphs correspond to $3$-connected hole-free 4-map graphs (see Section~\ref{sse:characterization}), and $n$-vertex hole-free 4-map graphs can be recognized in $O(n^3)$ time~\cite{ChenGP06}. See also~\cite{Brandenburg15}. Note that, since optimal $1$-planar graphs are triangulated, this implies that  a cubic time recognition algorithm exists also for these graphs.

\begin{itemize}

\item \cite{Brandenburg15}~F.~J.~Brandenburg.
\newblock On 4-map graphs and $1$-planar graphs and their recognition problem.
\newblock {\em CoRR}, abs/1509.03447, 2015.

\item \cite{ChenGP06}~Z.~Chen, M.~Grigni, and C.~H.~Papadimitriou.
\newblock Recognizing hole-free 4-map graphs in cubic time.
\newblock {\em Algorithmica}, 45(2):227--262, 2006.

\end{itemize}

Let $T=(V,E_T)$ be a maximal plane graph with $n$ vertices and let $M=(V,E_M)$ be a matching. Brandenburg \emph{et al.}~\cite{BrandenburgDEKL16} proved that there exists an $O(n^3)$-time algorithm to test if  $G=(V,E_T \cup E_M)$ admits an IC-planar drawing that preserves the embedding of $T$. If the test is positive, the algorithm computes a feasible drawing. The interest in this special case is  motivated by the fact that every IC-planar graph with maximum number of edges is the union of a triangulated planar graph and of a set of edges that form a matching~\cite{Zhang2013}.

\begin{itemize}

\item \cite{BrandenburgDEKL16}~F.~J.~Brandenburg, W.~Didimo, W.~S.~Evans, P.~Kindermann, G.~Liotta, and
  F.~Montecchiani.
\newblock Recognizing and drawing {IC}-planar graphs.
\newblock {\em Theor.~Comput.~Sci.}, 636:1--16, 2016.

\item \cite{Zhang2013}~X.~Zhang and G.~Liu.
\newblock The structure of plane graphs with independent crossings and its  applications to coloring problems.
\newblock {\em Open Math.}, 11(2):308--321, 2013.

\end{itemize}

Auer \emph{et al.}~\cite{AuerBBGHNR16} and Hong \emph{et al.}~\cite{HongEKLSS15} independently proved that recognizing outer $1$-planar graphs can be done efficiently. Both proofs are based on $O(n)$-time algorithms that decides whether an $n$-vertex input graph $G$ admits an outer $1$-planar embedding. In the positive case both algorithms return a valid embedding of $G$. In the negative case, the algorithm described in~\cite{AuerBBGHNR16} also returns one of six possible minors that are not outer $1$-planar.

\begin{itemize}

\item \cite{AuerBBGHNR16}~C.~Auer, C.~Bachmaier, F.~J.~Brandenburg, A.~Glei{\ss}ner, K.~Hanauer,
  D.~Neuwirth, and J.~Reislhuber.
\newblock Outer $1$-planar graphs.
\newblock {\em Algorithmica}, 74(4):1293--1320, 2016.

\item \cite{HongEKLSS15}~S.~Hong, P.~Eades, N.~Katoh, G.~Liotta, P.~Schweitzer, and Y.~Suzuki.
\newblock A linear-time algorithm for testing outer-$1$-planarity.
\newblock {\em Algorithmica}, 72(4):1033--1054, 2015.

\end{itemize}

Triangulated IC-planar and NIC-planar graphs can be recognized in cubic time~\cite{Brandenburg16b}, while optimal NIC-planar graphs can be recognized in linear time~\cite{BachmaierBHNR17}.

\begin{itemize}

\item \cite{BachmaierBHNR17}~C.~Bachmaier, F.~J.~Brandenburg, K.~Hanauer, D.~Neuwirth, and J.~Reislhuber.
\newblock NIC-planar graphs.
\newblock {\em CoRR}, abs/1701.04375, 2017.

\item \cite{Brandenburg16b}~F.~J.~Brandenburg.
\newblock Recognizing {IC}-planar and {NIC}-planar graphs.
\newblock {\em CoRR}, \\abs/1610.08884, 2016.

\end{itemize}

\subsubsection{Fixed rotation system setting}\label{ssse:fixed-rotation}
The recognition problem has also been studied with the additional assumption that the input graph comes along with a rotation system. Deciding whether a graph $G$ with a given rotation system $R$ admits a $1$-planar drawing that respects $R$ is NP-complete, even if $G$ is $3$-connected~\cite{AuerBGR15}. On the positive side, Eades \emph{et al.}~\cite{EHKLSS13} proved that there is an $O(n)$-time algorithm to decide if an $n$-vertex graph $G$ with a rotation system $R$ has a maximal $1$-planar embedding (i.e., a $1$-planar embedding in which no edge can be added without violating $1$-planarity) that respects $R$. In the positive case, the algorithm returns a valid embedding of $G$. The algorithm is based on the following two properties (proved in~\cite{EHKLSS13}). 
First, in any maximal $1$-planar embedding, the subgraph induced by the edges that do not intersect any other edge  is spanning and $2$-connected. Second, if $G$ admits a maximal $1$-planar embedding that respects $R$, then the embedding is unique.

\begin{itemize}

\item \cite{AuerBGR15}~C.~Auer, F.~J.~Brandenburg, A.~Glei{\ss}ner, and J.~Reislhuber.
\newblock $1$-planarity of graphs with a rotation system.
\newblock {\em J.~Graph Algorithms Appl.}, 19(1):67--86, 2015.

\item \cite{EHKLSS13}~P.~Eades, S.-H.~Hong, N.~Katoh, G.~Liotta, P.~Schweitzer, and Y.~Suzuki.
\newblock Testing maximal $1$-planarity of graphs with a rotation system in  linear time.
\newblock In {\em {GD} 2012}, volume 7704 of {\em LNCS}, pages 339--345.  Springer, 2013.

\end{itemize}

\subsubsection{Parameterized complexity}\label{ssse:parameterized}
Given its NP-hardness, it is natural to study the \oneplanarity in terms of parameterized complexity. That is, we seek  additional  parameters (other than the numbers of edges and vertices) that measure the complexity of an input graph, and  design recognition algorithms whose running time is the product of a polynomial in the input size and a non-polynomial function of these additional parameters. For more details on parameterized complexity theory, see~\cite{DowneyF99,FlumG06}. 
In this direction, Bannister {\em et al.}~\cite{BCE13} proved that, for an $n$-vertex graph $G$, deciding whether $G$ is $1$-planar is fixed parameter tractable for various graph parameters. More precisely, the problem can be solved in time $O(n + 2^{O(k^2)})$ if $G$ has cover number $k$; in time $O(n2^{2^{2d^2+O(d)}})$ if $G$ has tree-depth $d$; and in time $O(n + 2^{O((3c)!)})$ if $G$ has cyclomatic number $c$.

\begin{itemize}

\item \cite{BCE13}~M.~J.~Bannister, S.~Cabello, and D.~Eppstein.
\newblock Parameterized complexity of $1$-planarity.
\newblock In {\em {WADS} 2013}, volume 8037 of {\em LNCS}, pages 97--108.  Springer, 2013.
  
\item \cite{DowneyF99}~R.~G.~Downey and M.~R.~Fellows.
\newblock {\em Parameterized Complexity}.
\newblock Monographs in Computer Science. Springer, 1999.

\item \cite{FlumG06}~J.~Flum and M.~Grohe.
\newblock {\em Parameterized Complexity Theory}.
\newblock Texts in Theoretical Computer Science. An {EATCS} Series. Springer,  2006.

\end{itemize}

\section{Structural properties}\label{se:properties}

In this section we deal with structural properties and invariants of $1$-planar graphs. In Section~\ref{sse:coloring}, we review  results  on the classic coloring problem. In Section~\ref{sse:density}, we report bounds on the number of edges of $1$-planar graphs. Section~\ref{sse:edgepartitions} contains recent results on the problem of partitioning the edges of a $1$-planar graph, such that each partition set induces a graph with predefined properties. In Section~\ref{sse:parameters} we describe bounds on various graph parameters of $1$-planar graphs.  Section~\ref{sse:subgraphs} contains results about the existence of subgraphs of bounded vertex degree in $1$-planar graphs. Finally, results related with binary graph operations that preserve $1$-planarity are covered in Section~\ref{sse:operations}.

\subsection{Coloring}\label{sse:coloring}

A \emph{coloring} of a graph $G$ is an assignment of labels called \emph{colors} to elements of $G$ subject to certain constraints. Based on whether the elements of $G$ to be colored are its vertices, its edges, or both, we distinguish among vertex coloring, edge coloring, and total coloring, respectively. Coloring problems find application in task scheduling and frequency assignment
, register allocation
, as well as in pattern matching, designing seating plans, solving Sudoku puzzles and others
; see for example:

\begin{itemize}

\item \cite{Chaitin1982}~G.~J.~Chaitin.
\newblock Register allocation \& spilling via graph coloring.
\newblock {\em SIGPLAN Not.}, 17(6):98--101, 1982.

\item \cite{Lewis16}~R.~Lewis.
\newblock {\em A Guide to Graph Colouring: Algorithms and Applications}.
\newblock Springer, 2016.

\item \cite{Marx2004}~D.~Marx.
\newblock Graph colouring problems and their applications in scheduling.
\newblock {\em Periodica Polytechnica Ser.~El.~Eng.}, 48(1):11--16, 2004.

\end{itemize}

\subsubsection{Vertex coloring}

A \emph{proper vertex coloring} of a graph $G$ is an assignment of colors to the vertices of $G$ such that no two adjacent vertices receive the same color. The smallest number of colors needed to obtain a proper vertex coloring of a graph $G$ is called the \emph{chromatic number} of $G$.  A classic result in graph theory is that every planar graph has chromatic number at most $4$:

\begin{itemize}

\item \cite{MR1025335}~K.~Appel and W.~Haken.
\newblock {\em Every planar map is four colorable}, volume~98 of {\em  Contemporary Mathematics}.
\newblock AMS, 1989.

\end{itemize}

Ringel~\cite{R65} proved that every $1$-planar graph has chromatic number at most $7$, and conjectured that this bound could be lowered to $6$. Borodin~\cite{B84} settled the conjecture
in the affirmative, proving that the chromatic number of each $1$-planar graph is at most $6$. The bound is tight as for example the complete graph $K_6$ is $1$-planar and requires six colors. The same author later showed a relatively shorter proof of this result~\cite{B95}. However, Borodin's proof does not lead to an efficient algorithm for computing a proper vertex coloring with 6 colors of a $1$-planar graph. On the other hand, there is an $O(n)$-time algorithm for computing a proper vertex coloring with 7 colors of any $n$-vertex $1$-planar graph~\cite{ChenK05}. If we restrict to IC-planar graphs, then  Kr{\'{a}}l' and Stacho~\cite{KralS10} proved that the chromatic number of IC-planar graphs is $5$. Finally, it is  NP-complete to decide whether a given $1$-planar graph admits a proper vertex coloring using four colors~\cite{ChenK05}.

\begin{itemize}

\item \cite{B84}~O.~V.~Borodin.
\newblock Solution of the Ringel problem on vertex-face coloring of planar  graphs and coloring of $1$-planar graphs.
\newblock {\em Metody Diskret.~Analiz}, 108:12--26, 1984.

\item \cite{B95}~O.~V.~Borodin.
\newblock A new proof of the 6 color theorem.
\newblock {\em J.~Graph Theory}, 19(4):507--521, 1995.

\item \cite{ChenK05}~Z.~Chen and M.~Kouno.
\newblock A linear-time algorithm for 7-coloring $1$-plane graphs.
\newblock {\em Algorithmica}, 43(3):147--177, 2005.

\item \cite{KralS10}~D.~Kr{\'{a}}l' and L.~Stacho.
\newblock Coloring plane graphs with independent crossings.
\newblock {\em J.~Graph Theory}, 64(3):184--205, 2010.

\item \cite{R65}~G.~Ringel.
\newblock Ein {S}echsfarbenproblem auf der kugel.
\newblock {\em Abhandlungen aus dem Mathematischen Seminar der Universitaet  Hamburg}, 29(1--2):107--117, 1965.

\end{itemize}

Other forms of vertex coloring have been studied.
An \emph{acyclic coloring} of a graph $G$ is a proper vertex coloring of $G$ such that every cycle of $G$ uses at least three colors. The smallest number of colors needed to obtain an acyclic proper vertex coloring of a graph $G$ is called the \emph{acyclic chromatic number} of $G$. Borodin \emph{et al.}~\cite{BKRS01} proved that every $1$-planar graph has acyclic chromatic number at most $20$ and there is a $1$-planar graph which requires at least $7$ colors in any acyclic coloring.

\begin{itemize}

\item \cite{BKRS01}~O.~Borodin, A.~Kostochka, A.~Raspaud, and E.~Sopena.
\newblock Acyclic colouring of $1$-planar graphs.
\newblock {\em Discrete Appl.~Math.}, 114(1–3):29 -- 41, 2001.

\end{itemize}

Given a graph $G$ and a set $L(v)$ of colors for each vertex $v$ of $G$ (called a \emph{list}), a \emph{list coloring} is a proper vertex coloring of $G$ such that each vertex is assigned with a color in its corresponding list. Wang and Lih~\cite{WL08} proved that $1$-planar graphs have list colorings with at most seven colors.

\begin{itemize}

\item \cite{WL08}~W.~Wang and K.-W.~Lih.
\newblock Coupled choosability of plane graphs.
\newblock {\em J.~Graph Theory}, 58(1):27--44, 2008.

\end{itemize}

\subsubsection{Edge coloring}

A \emph{proper edge coloring} of a graph $G$ is an assignment of colors to the edges of $G$ such that no two adjacent edges receive the same color. The smallest number of colors needed to obtain a proper edge coloring of a graph $G$ is called the \emph{chromatic index} of $G$. By Vizing's theorem~\cite{Vizing64}, the number of colors needed to edge color a graph $G$ with maximum vertex degree $\Delta$ is either $\Delta$ or $\Delta+1$.

\begin{itemize}

\item \cite{Vizing64}~V.~G.~Vizing.
\newblock On an estimate of the chromatic class of a {$p$}-graph.
\newblock {\em Diskret.~Analiz No.}, 3:25--30, 1964.

\end{itemize}

Zhang and Wu~\cite{ZW11} proved that every $1$-planar graph with maximum vertex degree $\Delta \ge 10$ has chromatic index $\Delta$. Also, every $1$-planar graph without adjacent triangles and with maximum vertex degree $\Delta \ge 8$ has chromatic index $\Delta$~\cite{ZhangL2012}. Moreover, every $1$-planar graph without chordal 5-cycles and with maximum vertex degree $\Delta \ge 9$ has chromatic index $\Delta$, and for each $\Delta \le 7$ there exist $1$-planar graphs with maximum vertex degree $\Delta$ and chromatic index $\Delta+1$~\cite{ZhangL12}. The chromatic index of an IC-planar graph $G$ with maximum vertex degree $\Delta \ge 8$ is $\Delta$~\cite{Zhang2013}. The chromatic index of outer-$1$-planar graphs with maximum vertex
degree $\Delta \ge 4$ is $\Delta$, and there are infinitely many outer $1$-planar graphs with maximum vertex degree 3 and chromatic index 4~\cite{Zhang12}. Based on these results, Zhang~\cite{Zhang14d} argues that the chromatic index of any outer $1$-planar graph can be determined in polynomial time.

\begin{itemize}

\item \cite{Zhang14d}~X.~Zhang.
\newblock The edge chromatic number of outer-$1$-planar graphs.
\newblock {\em CoRR}, \\abs/1405.3183, 2014.

\item \cite{ZhangL2012}~X.~Zhang and G.~Liu.
\newblock On edge colorings of $1$-planar graphs without adjacent triangles.
\newblock {\em Inf.~Process.~Lett.}, 112(4):138 -- 142, 2012.

\item \cite{ZhangL12}~X.~Zhang and G.~Liu.
\newblock On edge colorings of $1$-planar graphs without chordal 5-cycles.
\newblock {\em Ars Comb.}, 104:431--436, 2012.

\item \cite{Zhang2013}~X.~Zhang and G.~Liu.
\newblock The structure of plane graphs with independent crossings and its  applications to coloring problems.
\newblock {\em Open Math.}, 11(2):308--321, 2013.

\item \cite{Zhang12}~X.~Zhang, G.~Liu, and J.-L.~Wu.
\newblock Edge covering pseudo-outerplanar graphs with forests.
\newblock {\em Discrete Math.}, 312(18):2788--2799, 2012.

\item \cite{ZW11}~X.~Zhang and J.-L.~Wu.
\newblock On edge colorings of $1$-planar graphs.
\newblock {\em Inf.~Process.~Lett.}, 111(3):124--128, 2011.
\end{itemize}

The list variant of the proper edge coloring problem has also been considered. Specifically, given a graph $G$ and a set $L(e)$ of colors for each edge $e$ of $G$ (called a \emph{list}), a \emph{list edge coloring} is a proper edge coloring of $G$ such that each edge is assigned with a color in its corresponding list. Every $1$-planar graph with maximum vertex degree $\Delta$ has a list edge coloring with $\Delta+1$ colors if $\Delta \ge 16$ and with $\Delta$ colors if $\Delta \ge 21$~\cite{ZWL12}. If $G$ is IC-planar and $\Delta \ge 8$, then it has a list edge coloring with $\Delta$ colors~\cite{Zhang2013}.

\begin{itemize}

\item \cite{Zhang2013}~X.~Zhang and G.~Liu.
\newblock The structure of plane graphs with independent crossings and its  applications to coloring problems.
\newblock {\em Open Math.}, 11(2):308--321, 2013.

\item \cite{ZWL12}~X.~Zhang, J.~Wu, and G.~Liu.
\newblock List edge and list total coloring of $1$-planar graphs.
\newblock {\em Front.~Math. China}, 7(5):1005--1018, 2012.

\end{itemize}

\subsubsection{Total coloring}

A \emph{total coloring} of a graph $G$ is an assignment of colors to the vertices and edges of $G$ such that neither two adjacent edges, two adjacent vertices nor an edge and its end-vertices  receive the same color. The smallest number of colors needed to obtain a total coloring of a graph $G$ is called the \emph{total chromatic number} of $G$. Behzad~\cite{Behzad65} conjectured that the total chromatic number of a graph $G$ with maximum vertex degree $\Delta$ is at most $\Delta+2$.

\begin{itemize}

\item \cite{Behzad65}~M.~Behzad.
\newblock {\em Graphs and Their Chromatic Numbers}.
\newblock PhD thesis, Michigan State University, Department of Mathematics,  1965.

\end{itemize}

Zhang {\em et al.}~\cite{ZhangHL15} proved that the total coloring conjecture holds for $1$-planar graphs with maximum vertex degree $\Delta \ge 13$. Also, Czap~\cite{C13} proved that the total chromatic number of a $1$-planar graph $G$ with maximum vertex degree $\Delta \ge 10$  is at most $\Delta+3$.

\begin{itemize}

\item \cite{C13}~J.~Czap.
\newblock A note on total colorings of $1$-planar graphs.
\newblock {\em Inf.~Process.~Lett.}, 113(14-16):516--517, 2013.

\item \cite{ZhangHL15}~X.~Zhang, J.~Hou, and G.~Liu.
\newblock On total colorings of $1$-planar graphs.
\newblock {\em J.~Comb.~Optim.}, 30(1):160--173, 2015.

\end{itemize}

The list variant of the total coloring problem has also been studied. Specifically, given a graph $G$ and a set 
$L(e)$ (respectively, $L(v)$) of colors for each edge $e$ of $G$ (respectively, vertex $v$ of $G$), a \emph{list total coloring} is a total coloring of $G$ such that each edge and each vertex is assigned with a color in its corresponding list. Every $1$-planar graph with maximum vertex degree $\Delta$ has a list total coloring with $\Delta+2$ colors if $\Delta \ge 16$ and with $\Delta+1$ colors if $\Delta \ge 21$~\cite{ZWL12}. If $G$ is IC-planar and $\Delta \ge 8$, then it has a list total coloring with $\Delta+1$ colors~\cite{Zhang2013}.

\begin{itemize}

\item \cite{Zhang2013}~X.~Zhang and G.~Liu.
\newblock The structure of plane graphs with independent crossings and its  applications to coloring problems.
\newblock {\em Open Math.}, 11(2):308--321, 2013.

\item \cite{ZWL12}~X.~Zhang, J.~Wu, and G.~Liu.
\newblock List edge and list total coloring of $1$-planar graphs.
\newblock {\em Front.~Math. China}, 7(5):1005--1018, 2012.

\end{itemize}

\subsection{Edge density}\label{sse:density}

The fact that a $1$-planar graph $G$ with $n$ vertices has at most $4n-8$ edges has been proved multiple times~\cite{BSW83,ChenZ07,FM07,PachT97}. This bound is shown to be tight for every $n = 8$ and for $n \ge 10$~\cite{BSW83}. Given the characterization of optimal $1$-planar graphs discussed in Section~\ref{sse:characterization}, this result can be easily observed as follows. Let $G$ be an optimal $1$-planar graph with $n$ vertices, and consider the underlying $3$-connected planar quadrangulation $Q$ of $G$. Graph $Q$ has $n-2$ faces and $2n-4$ edges. Inserting a pair of crossing edges in each face of $Q$ adds $2(n-2)$ edges, which leads to $4n-8$ edges in total. An optimal $1$-planar graph is shown in Figure~\ref{fi:optimal1planar}.

\begin{itemize}

\item \cite{BSW83}~R.~Bodendiek, H.~Schumacher, and K.~Wagner.
\newblock Bemerkungen zu einem {S}echsfarbenproblem von G.~Ringel.
\newblock {\em Abhandlungen aus dem Mathematischen Seminar der Universitaet Hamburg}, 53(1):41--52, 1983.

\item \cite{ChenZ07}~Z.-Z.~Chen.
\newblock New bounds on the edge number of a $k$-map graph.
\newblock {\em J.~Graph Theory}, 55(4):267--290, 2007.

\item \cite{FM07}~I.~Fabrici and T.~Madaras.
\newblock The structure of $1$-planar graphs.
\newblock {\em Discrete Math.}, 307(7--8):854--865, 2007.

\item \cite{PachT97}~J.~Pach and G.~T{\'o}th.
\newblock Graphs drawn with few crossings per edge.
\newblock {\em Combinatorica}, 17(3):427--439, 1997.

\end{itemize}

If we consider only those $1$-planar graphs that admit a $1$-planar straight-line drawing (see Section~\ref{se:geometric-representations}), then any such graph with $n$-vertices has at most $4n-9$ edges, as proved by Didimo~\cite{D13} (see also the alternative proof by Ackerman~\cite{Ackerman14}).

\begin{itemize}

\item \cite{Ackerman14}~E.~Ackerman.
\newblock A note on $1$-planar graphs.
\newblock {\em Discrete Appl.~Math.}, 175:104--108, 2014.

\item \cite{D13}~W.~Didimo.
\newblock Density of straight-line $1$-planar graph drawings.
\newblock {\em Inf.~Process.~Lett.}, 113(7):236--240, 2013.

\end{itemize}

\begin{figure}
\centering
\subfigure[]{\includegraphics[width=0.32\columnwidth,page=1]{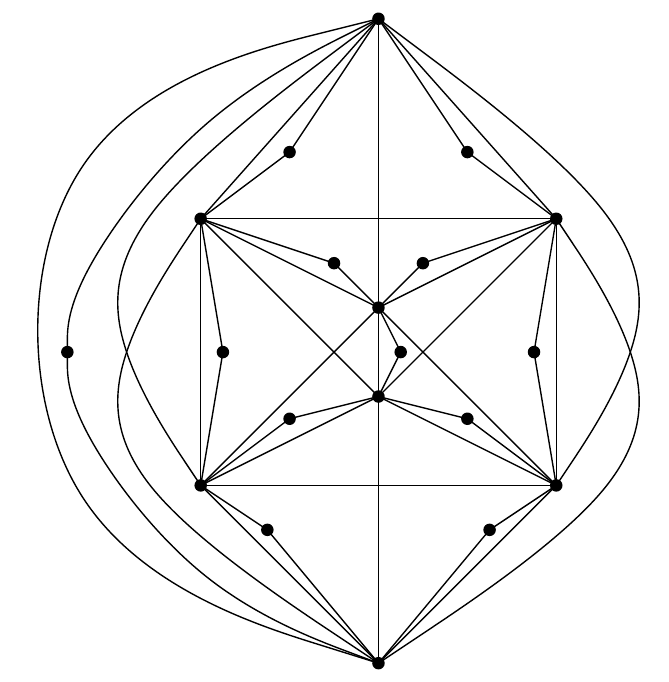}\label{fi:density-maximal}}\hfil
\subfigure[]{\includegraphics[width=0.32\columnwidth,page=2]{figs/density}\label{fi:density-ic}}\hfil
\subfigure[]{\includegraphics[width=0.32\columnwidth,page=3]{figs/density}\label{fi:density-nic}}
\caption{(a) A maximal $1$-planar graph with 20 vertices and 44 edges. (b) A maximal IC-planar graph with 8 vertices and 20 edges. (c) A maximal NIC-planar graph with 12 vertices and 36 edges.}
\end{figure}

Density considerations of $1$-planar graphs are fundamentally different from the equivalent considerations for planar graphs. For example, 
Brandenburg \emph{et al.}~\cite{BEGGHR13} proved that there exist  maximal $1$-planar graphs with $n$ vertices and $\frac{45}{17}n-\frac{84}{17}<2.65n$ edges; see Figure~\ref{fi:density-maximal} for a sparse maximal graph obtained with the construction in~\cite{BEGGHR13}. 
This is in contrast with maximal planar graphs, which have exactly $3n-6$ edges. Moreover, any $n$-vertex maximal $1$-planar graph has at least $\frac{28}{13}n-\frac{10}{3}$ edges~\cite{BEGGHR13}. Brandenburg \emph{et al.}~\cite{BEGGHR13} also studied the problem in the fixed rotation system setting. They proved that, for infinitely many values of $n$, there are $n$-vertex maximal $1$-planar graphs with a fixed rotation system that have $\frac{7}{3}n-3$ edges, and that any maximal $1$-planar graph with a fixed rotation system has at least $\frac{21}{10}n-\frac{10}{3}$ edges.

\begin{itemize}

\item \cite{BEGGHR13}~F.-J.~Brandenburg, D.~Eppstein, A.~Glei{\ss}ner, M.~T.~Goodrich, K.~Hanauer,
  and J.~Reislhuber.
\newblock On the density of maximal $1$-planar graphs.
\newblock In {\em {GD} 2012}, volume 7704 of {\em LNCS}, pages 327--338.  Springer, 2013.

\end{itemize}

An IC-planar graph $G$ with $n$ vertices has at most $3.25n - 6$ edges and this bound is tight~\cite{Zhang2013}. The graph in Figure~\ref{fi:density-ic} is used in~\cite{Zhang2013} to prove the tightness of the result. A NIC-planar graph $G$ with $n$ vertices has at most $3.6n - 7.2$ edges and this bound is tight~\cite{CzapS17} (see also~\cite{BachmaierBHNR17}). For example, Figure~\ref{fi:density-nic} shows a maximal NIC-planar graph that matches this bound.

\begin{itemize}

\item \cite{BachmaierBHNR17}~C.~Bachmaier, F.~J.~Brandenburg, K.~Hanauer, D.~Neuwirth, and J.~Reislhuber.
\newblock NIC-planar graphs.
\newblock {\em CoRR}, abs/1701.04375, 2017.

\item \cite{CzapS17}~J.~{Czap} and P.~{{\v S}ugerek}.
\newblock Drawing graph joins in the plane with restrictions on crossings.
\newblock {\em Filomat}, 31(2):363–--370, 2017.

\item \cite{Zhang2013}~X.~Zhang and G.~Liu.
\newblock The structure of plane graphs with independent crossings and its  applications to coloring problems.
\newblock {\em Open Math.}, 11(2):308--321, 2013.

\end{itemize}

\begin{figure}
\centering
\subfigure[]{\includegraphics[width=0.32\columnwidth,page=4]{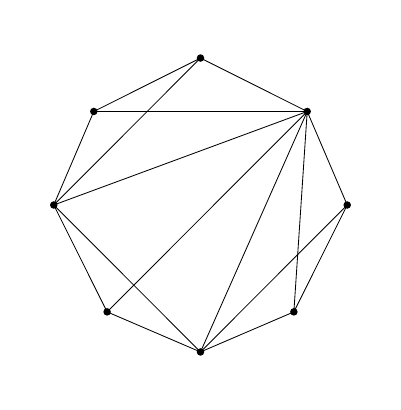}\label{fi:density-o1p-1}}\hfil
\subfigure[]{\includegraphics[width=0.32\columnwidth,page=5]{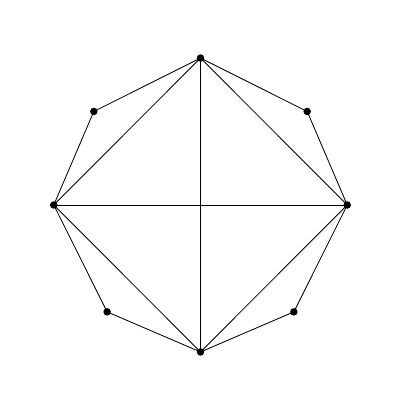}\label{fi:density-o1p-2}}\hfil
\subfigure[]{\includegraphics[width=0.32\columnwidth,page=6]{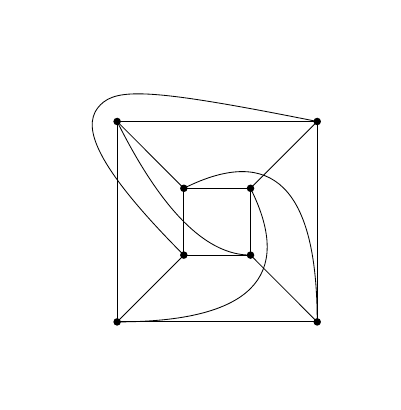}\label{fi:density-bipartite}}
\caption{(a) A maximal outer $1$-planar graph with 8 vertices and 16 edges. (b) A maximal outer $1$-planar graph with 8 vertices and 14 edges. (c) A maximal bipartite $1$-planar graph with 8 vertices and 16 edges. }
\end{figure}

Auer \emph{et al.}~\cite{AuerBBGHNR16} proved that an outer $1$-planar graph $G$ with $n$ vertices has at most $2.5n-4$ edges and this bound is tight. Furthermore, every maximal outer $1$-planar graph with $n$ vertices has at least $2.2n - 3.6$ edges and there exist maximal outer $1$-planar graphs that match this bound~\cite{AuerBBGHNR16}.
\begin{itemize}

\item \cite{AuerBBGHNR16}~C.~Auer, C.~Bachmaier, F.~J.~Brandenburg, A.~Glei{\ss}ner, K.~Hanauer,
  D.~Neuwirth, and J.~Reislhuber.
\newblock Outer $1$-planar graphs.
\newblock {\em Algorithmica}, 74(4):1293--1320, 2016.

\end{itemize}

Karpov~\cite{Karpov2014} proved that a bipartite $1$-planar graph with $n$ vertices has at most $3n - 8$ edges for even $n$ such that $n \neq 6$, and at most $3n - 9$ edges for odd $n$ and for $n = 6$. Also, for all $n  \ge  4$, there exist examples showing that these bounds are tight~\cite{Karpov2014}; Figure~\ref{fi:density-bipartite} shows one of these graphs. Czap \emph{et al.}~\cite{CzapPS16} showed that the maximum number of edges of bipartite $1$-planar graphs that are almost balanced is not significantly smaller than $3n - 8$, while the same is not true for unbalanced ones.  In particular, if the size of the smaller partite set is sublinear in $n$, then the number of edges is $(2 + o(1))n$, while the same is not true otherwise.

\begin{itemize}

\item \cite{CzapPS16}~J.~Czap, J.~Przybylo, and E.~Skrabul'{\'{a}}kov{\'{a}}.
\newblock On an extremal problem in the class of bipartite 1-planar graphs.
\newblock {\em Discuss. Math. Graph Theory}, 36(1):141--151, 2016.

\item \cite{Karpov2014}~D.~V.~Karpov.
\newblock An upper bound on the number of edges in an almost planar bipartite
  graph.
\newblock {\em J.~Math.~Sci.}, 196(6):737--746, 2014.

\end{itemize}

\subsection{Edge partitions}\label{sse:edgepartitions}

A well-studied subject in graph algorithms and graph theory is the coloring of the edges of a graph such that each monochromatic set induces a subgraph with special properties. This problem is studied in great details for planar graphs; see for example:

\begin{itemize}

\item \cite{Ding2000221}~G.~Ding, B.~Oporowski, D.~P.~Sanders, and D.~Vertigan.
\newblock Surfaces, tree-width, clique-minors, and partitions.
\newblock {\em J.~Combin.~Theory Ser.~B}, 79(2):221 -- 246, 2000.

\item \cite{ec-pepg+-88}~E.~S.~Elmallah and C.~J.~Colbourn.
\newblock Partitioning the edges of a planar graph into two partial $k$-trees.
\newblock In {\em {CGTC} 1988}, Congressus numerantium, pages 69--80. Utilitas  Mathematica, 1988.

\item \cite{Goncalves05}~D.~Gon{\c{c}}alves.
\newblock Edge partition of planar graphs into two outerplanar graphs.
\newblock In {\em {STOC} 2005}, pages 504--512. {ACM}, 2005.

\item \cite{Kedlaya1996238}~K.~S.~Kedlaya.
\newblock Outerplanar partitions of planar graphs.
\newblock {\em J.~Combin.~Theory Ser.~B}, 67(2):238 -- 248, 1996.

\end{itemize}

In the context of $1$-planar graphs, an \emph{edge partition} of a $1$-planar graph $G$ is an edge coloring of $G$ with two colors, say \emph{red} and \emph{blue}, such that both the graph formed by the red edges, called the \emph{red graph}, and the graph formed by the blue edges, called the \emph{blue graph}, are planar. Note that, given a $1$-planar embedding of $G$, an edge partition of $G$ can be constructed by coloring red an edge for each pair of crossing edges, and by coloring blue the remaining edges.
Czap and Hud{\'a}k~\cite{CH13} proved that every optimal $1$-planar graph admits an edge partition such that the red graph is a forest. This result has been later extended to all $1$-planar graphs by Ackerman~\cite{Ackerman14}.

\begin{itemize}

\item \cite{Ackerman14}~E.~Ackerman.
\newblock A note on $1$-planar graphs.
\newblock {\em Discrete Appl.~Math.}, 175:104--108, 2014.

\item \cite{CH13}~J.~Czap and D.~Hud{\'a}k.
\newblock On drawings and decompositions of $1$-planar graphs.
\newblock {\em Electr. J.~Comb.}, 20(2):P54, 2013.

\end{itemize}

Motivated by visibility representations of $1$-planar graphs (see also Section~\ref{sse:visibility}), Lenhart \emph{et al.}~\cite{LenhartLM17} and Di Giacomo \emph{et al.}~\cite{DiGiacomo2017} studied edge partitions such that the red graph has maximum vertex degree that is bounded by a constant independent of the size of the graph. Specifically, Lenhart \emph{et al.}~\cite{LenhartLM17} proved that if $G$ is an $n$-vertex optimal $1$-plane graph with an edge partition with the red graph $G_R$ being a forest, then $G_R$ has $n$ vertices (i.e., it is a spanning subgraph of $G$) and it is composed of two trees. Based on this finding, they  proved that for any constant $c$, there exists an optimal $1$-plane graph such that in any edge partition with the red graph being a forest, the maximum vertex degree of the red graph is at least $c$. On the positive side, if we drop the acyclicity requirement, then every  optimal $1$-planar graph admits an edge partition such that the red graph has maximum vertex degree $4$, and degree $4$ is sometimes needed~\cite{LenhartLM17}. Also, every $3$-connected $1$-planar graph admits an edge partition such that the red graph has maximum vertex degree six, and degree six is sometimes needed, as shown by Di Giacomo \emph{et al.}~\cite{DiGiacomo2017}. Finally, for every $n>0$ there exists an $O(n)$-vertex 2-connected $1$-planar graph such that in any edge partition the red graph has maximum vertex degree $\Omega(n)$~\cite{DiGiacomo2017}.

\begin{itemize}

\item \cite{DiGiacomo2017}~E.~{Di Giacomo}, W.~Didimo, W.~S. Evans, G.~Liotta, H.~Meijer, F.~Montecchiani, and S.~K. Wismath.
\newblock Ortho-polygon visibility representations of embedded graphs.
\newblock {\em Algorithmica}, 2017.

\item \cite{LenhartLM17}~W.~J.~Lenhart, G.~Liotta, and F.~Montecchiani.
\newblock On partitioning the edges of $1$-plane graphs.
\newblock {\em Theor.~Comput.~Sci.}, 662:59--65, 2017.

\end{itemize}

More recently, Di Giacomo \emph{et al.}~\cite{DiGiacomoEtAl2017} proved that every NIC-plane graph admits an edge partition such that the red graph has maximum vertex degree three, and that this bound on the vertex degree is worst-case optimal. Moreover, deciding whether a $1$-plane graph admits an edge partition such that the red graph has maximum vertex degree two is NP-complete. On the positive side, deciding whether an $n$-vertex $1$-plane graph admits an edge partition such that the red graph has maximum vertex degree one, and computing one in the positive case, can be done in $O(n^2)$ time.

\begin{itemize}

\item \cite{DiGiacomoEtAl2017}~E.~{Di Giacomo}, W.~Didimo, W.~S. Evans, G.~Liotta, H.~Meijer, F.~Montecchiani, and S.~K. Wismath.
\newblock New results on edge partitions of 1-plane graphs.
\newblock {\em CoRR}, abs/1706.05161, 2017.

\end{itemize}

\subsection{Graph parameters}\label{sse:parameters}

In this section we present results that deal with bounds on various graph parameters for the family of $1$-planar graphs. In particular, we first present recent results on the book thickness of $1$-planar graphs (Section~\ref{ssse:thickness}), and then list some bounds on the treewidth  (Subsection~\ref{ssse:treewidth}) and on the expansion (Subsection~\ref{ssse:expansion}) of these graphs.

\begin{figure}
\centering
\subfigure[]{\includegraphics[width=0.45\columnwidth,page=1]{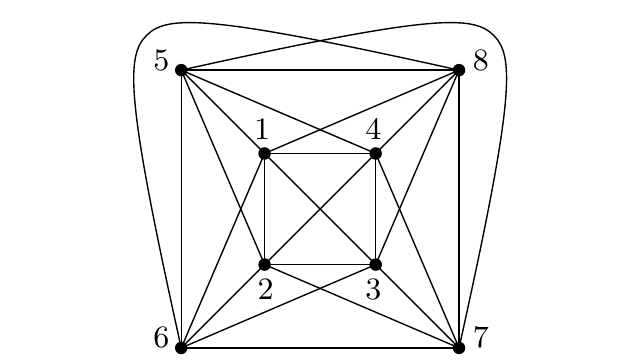}\label{fi:bt-graph}}
\subfigure[]{\includegraphics[width=0.45\columnwidth,page=2]{figs/bookthickness}\label{fi:bt-4pages}}
\caption{(a) A $1$-planar graph $G$ with book thickness $4$. (b) A 4-page book embedding of $G$. Solid edges above (below) the spine belong to the same page, as well as dotted edges above (below) the spine.\label{fi:bt}}
\end{figure}

\subsubsection{Book thickness} \label{ssse:thickness}

A \emph{$k$-page book embedding}, for some integer $k \geq 0$, is a particular representation of a graph $G$ with the following properties: $(i)$ The vertices are restricted to a line, called the \emph{spine}; $(ii)$ The edges are partitioned into $k$ sets, called \emph{pages}, such that edges in a same page are drawn on a half-plane delimited by the spine and do not cross each other. The minimum $k$ such that $G$ has a $k$-page book embedding is the \emph{book thickness} of $G$ (also known as \emph{pagenumber} and \emph{stacknumber}). Book embeddings have applications in VLSI design
, stack sorting
, traffic control
, graph drawing
, and more; see for example:

\begin{itemize}

\item \cite{Angelini2012}~P.~Angelini, G.~{Di Battista}, F.~Frati, M.~Patrignani, and I.~Rutter.
\newblock Testing the simultaneous embeddability of two graphs whose  intersection is a $2$-connected or a connected graph.
\newblock {\em J.~Discrete Algorithms}, 14:150 -- 172, 2012.

\item \cite{Baur2005}~M.~Baur and U.~Brandes.
\newblock Crossing reduction in circular layouts.
\newblock In {\em {WG} 2004}, volume 3353 of {\em LNCS}, pages 332--343.  Springer, 2005.

\item \cite{ChungLR1987}~F.~R.~K. Chung, F.~T.~Leighton, and A.~L.~Rosenberg.
\newblock Embedding graphs in books: A layout problem with applications to {VLSI} design.
\newblock {\em {SIAM} J.~Algebraic Discr.~Meth.}, 8(1):33--58, 1987.

\item \cite{Kainen90}~P.~C.~Kainen.
\newblock The book thickness of a graph.~{II}.
\newblock In {\em {CGTC} 1990}, number v.~5 in Congressus numerantium, pages  127--132. Utilitas Mathematica, 1990.

\item \cite{McKenzieO10}~T.~McKenzie and S.~Overbay.
\newblock Book embeddings and zero divisors.
\newblock {\em Ars Comb.}, 95, 2010.

\item \cite{Wattenberg2002}~M.~Wattenberg.
\newblock Arc diagrams: visualizing structure in strings.
\newblock In {\em IEEE {INFOVIS} 2002.}, pages 110--116. IEEE, 2002.

\item \cite{Wood2002}~D.~R.~Wood.
\newblock Bounded degree book embeddings and three-dimensional orthogonal graph  drawing.
\newblock In {\em {GD} 2001}, volume 2265 of {\em LNCS}, pages 312--327.  Springer, 2002.

\end{itemize}

One of the fundamental results is that the book thickness of planar graphs is at most $4$~\cite{Yannakakis1989}, although a planar graph requiring four pages is yet to be found.

\begin{itemize}

\item \cite{Yannakakis1989}~M.~Yannakakis.
\newblock Embedding planar graphs in four pages.
\newblock {\em J.~Comput.~Syst.~Sci.}, 38(1):36 -- 67, 1989.

\end{itemize}

Since a graph with $m$ edges has book thickness $O(\sqrt{m})$~\cite{Malitz1994}, it immediately follows that an $n$-vertex $1$-planar graph has book thickness $O(\sqrt{n})$ (see also Section~\ref{sse:density}). A constant upper bound equal to 39 for the book thickness of $1$-planar graphs has been proved by Bekos \emph{et al.}~\cite{BekosB0R15}. Alam \emph{et al.}~\cite{AlamBK15} further improved this bound to 16 for general $1$-planar graphs, and to 12 for $3$-connected $1$-planar graphs. Bekos \emph{et al.}~\cite{Bekos0Z15} observed that there are $1$-planar graphs with book thickness exactly $4$; see also Figure~\ref{fi:bt}. A SAT formulation for the book thickness problem has been given in the same paper~\cite{Bekos0Z15}, where the authors also hypothesized the existence of $1$-planar graphs requiring at least five pages, although the experiments could not confirm this hypothesis.

\begin{itemize}

\item \cite{AlamBK15}~M.~J.~Alam, F.~J.~Brandenburg, and S.~G.~Kobourov.
\newblock On the book thickness of $1$-planar graphs.
\newblock {\em CoRR}, abs/1510.05891, 2015.

\item \cite{BekosB0R15}~M.~A.~Bekos, T.~Bruckdorfer, M.~Kaufmann, and C.~N.~Raftopoulou.
\newblock $1$-planar graphs have constant book thickness.
\newblock In {\em {ESA} 2015}, volume 9294 of {\em LNCS}, pages 130--141.  Springer, 2015.

\item \cite{Bekos0Z15}~M.~A.~Bekos, M.~Kaufmann, and C.~Zielke.
\newblock The book embedding problem from a {SAT}-solving perspective.
\newblock In {\em {GD} 2015}, volume 9411 of {\em LNCS}, pages 125--138.  Springer, 2015.

\item \cite{Malitz1994}~S.~Malitz.
\newblock Graphs with {$E$} edges have pagenumber {$O(\sqrt{E})$}.
\newblock {\em J.~Algorithms}, 17(1):71 -- 84, 1994.

\end{itemize}

\subsubsection{Treewidth}\label{ssse:treewidth} 
A \emph{tree decomposition} of a graph $G=(V,E)$ is a tree $T$ and a one-to-one mapping from the vertex set of $T$ to a collection $B_1,\dots,B_k$ of subsets of $V$ such that: $(i)$ every vertex of $G$ is in $B_i$ for some $1 \leq i \leq k$; $(ii)$ every edge of $G$ has both end-vertices in $B_i$ for some $1 \leq i \leq k$; $(iii)$ If $B_i$ and $B_j$ ($i \neq j$) both contain a vertex $v$ of $G$, then all vertices of $T$ in the (unique) path between $B_i$ and $B_j$ contain $v$ as well. The \emph{width} of a decomposition is one less than the maximum number of vertices in any subset $B_i$. The \emph{treewidth} of a graph is the minimum width of any tree decomposition. The \emph{pathwidth} of a graph can be defined analogously as  the treewidth but is restricted to tree decompositions in which the tree $T$ is a path.

Dujmovi{\'c} \emph{et al.}~\cite{2015arXiv150604380D} proved that an $n$-vertex $1$-planar graph $G$ has pathwidth and  treewidth at most $O(\sqrt{n})$. This upper bound also implies that $G$ has a $\frac{1}{2}$-separator of size at most $O(\sqrt{n})$. We recall that for $\varepsilon \in (0, 1)$, a set $S$ of vertices in an $n$-vertex graph $G$ is an $\varepsilon$-separator of $G$ if each component of $G \setminus S$ has at most $\varepsilon \cdot n$ vertices. A similar upper bound on the size of balanced separators of $1$-planar graphs was already proved by Grigoriev and Boadlander~\cite{GB07}, who took advantage of this result to show that many optimization problems admit polynomial time approximation schemes (PTAS) when restricted to these graphs~\cite{GB07}.

In addition, Dujmovi{\'c} \emph{et al.}~\cite{2015arXiv150604380D} proved that every $1$-planar graph has layered treewidth at most 12 (see~\cite{DujimovicMW2017} for the definition of layered treewidth).

\begin{itemize}

\item \cite{2015arXiv150604380D}~V.~{Dujmovi{\'c}}, D.~{Eppstein}, and D.~R. {Wood}.
\newblock Structure of graphs with locally restricted crossings.
\newblock {\em SIAM J. Discrete Math.}, 31(2):805--824, 2017.

\item \cite{DujimovicMW2017}~V.~{Dujmovi{\'c}}, P.~{Morin}, and D.~R. {Wood}.
\newblock {Layered separators in minor-closed families with applications}.
\newblock {\em J. Combin. Theory Ser. B.}, 2017.
\newblock doi:10.1016/j.jctb.2017.05.006.

\item \cite{GB07}~A.~Grigoriev and H.~L.~Bodlaender.
\newblock Algorithms for graphs embeddable with few crossings per edge.
\newblock {\em Algorithmica}, 49(1):1--11, 2007.

\end{itemize}

\subsubsection{Expansion}\label{ssse:expansion}
A \emph{$t$-shallow minor} of a graph $G$ is a graph formed from $G$ by contracting a collection of vertex-disjoint subgraphs of radius $t$, and deleting the remaining vertices of $G$. A family of graphs has \emph{bounded expansion} if there exists a function $f$ such that, in every $t$-shallow minor of a graph in the family, the number of edges over the number vertices is at most $f(t)$.  Graphs with bounded expansion have interesting properties as shown in~\cite{DBLP:books/daglib/0030491}, and if the function $f$ is polynomial than there exist polynomial-time approximation schemes for several  optimization problem~\cite{Har-PeledQ15}. Note that the graphs with bounded expansion form a broader class than those that are minor-closed.

Ne{\v s}et{\v r}il \emph{et al.}~\cite{NesetrilMW12} proved that $1$-planar graphs have bounded expansion. This is also implied by the fact that $n$-vertex $1$-planar graphs have $\frac{1}{2}$-separators of size at most $O(\sqrt{n})$~\cite{2015arXiv150604380D}.

\begin{itemize}

\item \cite{Har-PeledQ15}~S.~Har{-}Peled and K.~Quanrud.
\newblock Approximation algorithms for polynomial-expansion and low-density  graphs.
\newblock In {\em {ESA} 2015}, volume 9294 of {\em LNCS}, pages 717--728. Springer, 2015.

\item \cite{DBLP:books/daglib/0030491}~J.~Ne{\v s}et{\v r}il and P.~O. de~Mendez.
\newblock {\em Sparsity - Graphs, Structures, and Algorithms}, volume~28 of  {\em Algorithms and combinatorics}.
\newblock Springer, 2012.

\item \cite{NesetrilMW12}~J.~Ne{\v s}et{\v r}il, P.~O. de~Mendez, and D.~R. Wood.
\newblock Characterisations and examples of graph classes with bounded
  expansion.\newblock {\em Eur.~J.~Comb.}, 33(3):350--373, 2012.

\end{itemize}

\subsection{Subgraphs with bounded vertex degree}\label{sse:subgraphs}

In this section we review some results concerning the existence of subgraphs of bounded vertex degree in $1$-planar graphs.

Fabrici and Madaras~\cite{FM07} proved that each $1$-planar graph contains a vertex
of degree at most 7, and that each $3$-connected $1$-planar graph contains an edge with both end-vertices of degree at most 20. Hud{\'{a}}k and {\v S}ugerek~\cite{HudakS12} proved that each $1$-planar graph of minimum degree $\Delta \ge 4$ contains an edge having one end-vertex with degree $4$ and the other with degree at most $13$, or one with degree $5$ and the other one with degree  at most $9$, or one with degree $6$ and the other one with degree at most $8$, or both with degree $7$. 

\begin{itemize}

\item \cite{FM07}~I.~Fabrici and T.~Madaras.
\newblock The structure of $1$-planar graphs.
\newblock {\em Discrete Math.}, 307(7--8):854--865, 2007.

\item \cite{HudakS12}~D.~Hud{\'{a}}k and P.~{\v S}ugerek.
\newblock Light edges in $1$-planar graphs with prescribed minimum degree.
\newblock {\em Discuss. Math. Graph Theory}, 32(3):545--556, 2012.

\end{itemize}

Hud{\'a}k and Madaras~\cite{HM09} proved that each $1$-planar graph of minimum vertex degree 5 and girth\footnote{The girth of a graph is the length of the shortest cycles contained in the graph.} 4 contains: $(i)$ a vertex with degree 5 adjacent to a vertex with degree at most 6; $(ii)$ a 4-cycle such that every vertex in this cycle has degree at most 9, $(iii)$ a complete bipartite graph $K_{1,4}$ with all vertices having degree at most 11.

\begin{itemize}

\item \cite{HM09}~D.~Hud{\'a}k and T.~Madaras.
\newblock On local structure of $1$-planar graphs of minimum degree 5 and girth  4.
\newblock {\em Discuss.~Math.~Graph Theory}, 29(2):385--400, 2009.

\end{itemize}

Hud{\'a}k \emph{et al.}~\cite{HMS12} proved that every optimal $1$-planar graphs with at least $k$ vertices contains a path on $k$ vertices such that the sum of the degrees of these $k$ vertices is at most $8k - 1$; and that each $3$-connected maximal $1$-planar graph with at least $2k$ vertices contains a path on $k$ vertices whose vertices have degree at most $10k$ each.

\begin{itemize}

\item \cite{HMS12}~D.~Hud{\'a}k, T.~Madaras, and Y.~Suzuki.
\newblock On properties of maximal $1$-planar graphs.
\newblock {\em Discuss.~Math.~Graph Theory}, 32(4):737--747, 2012.

\end{itemize}

\subsection{Binary operations}\label{sse:operations}

In this section we report results concerning binary graph operations that preserve $1$-planarity.

The \emph{join product} $G + H$ of two graphs $G$ and $H$ is obtained from vertex-disjoint copies of $G$ and $H$ by adding all edges between the vertices of $G$ and the vertices of $H$.
Let $C_n$ and $P_n$ denote the cycle and the path with $n$ vertices, respectively. Czap \emph{et al.}~\cite{Czap2014} studied the problem of characterizing those graph pairs whose join product yields a $1$-planar graph. They proved that, in the case when both $G$ and $H$ have at least three vertices, the join $G + H$ is $1$-planar if and only if the pair $[G, H]$ is subgraph-majorized (that is, both $G$ and $H$ are subgraphs of graphs of the major pair) by one of pairs $[C_3 \cup C_3, C_3]$, $[C_4, C_4]$, $[C_4, C_3]$, $[K_{2,1,1}, P_3]$.

\begin{itemize}

\item \cite{Czap2014}
J.~Czap, D.~Hud{\'a}k, and T.~Madaras.
\newblock Joins of $1$-planar graphs.
\newblock {\em Acta Math.~Sin.~English Series}, 30(11):1867--1876, 2014.

\end{itemize}

A \emph{lexicographic product} $G \circ H$ of two graphs $G$ and $H$ is a graph whose vertex set is the Cartesian product of the vertex set of $G$ and the vertex set of $H$, and two vertices $(u, v)$ and $(x, y)$ are adjacent in $G \circ H$ if and only if either $u$ is adjacent to $x$ in $G$ or $u = x$ and $v$ is adjacent to $y$ in $H$. Note that the lexicographic product is not commutative, i.e. $G \circ H \neq H \circ G$ in general. Bucko and Czap~\cite{Bucko2015} studied lexicographic products that yield a $1$-planar graph. For example, they proved that if $G$ has minimum vertex degree three and $H$ is a single vertex, then $G \circ H$ is $1$-planar if and only if $G$ is $1$-planar.

\begin{itemize}

\item \cite{Bucko2015}
J.~Bucko and J.~Czap.
\newblock $1$-planar lexicographic products of graphs.
\newblock {\em Applied Mathematical Sciences}, 9(109):5441--5449, 2015.

\end{itemize}

\section{Geometric representations}\label{se:geometric-representations}

This section is devoted to geometric representations of $1$-planar graphs. In Section~\ref{sse:straightline}, we survey results on the problem of computing straight-line drawings of  $1$-planar graphs. In Section~\ref{sse:rac}, we report results on straight-line and polyline drawings of $1$-planar graphs with the additional property that edges cross only at right angles. Finally, Sections~\ref{sse:visibility} and~\ref{sse:contact} contain results on visibility representations and contact representations of $1$-planar graphs, respectively.

\subsection{Straight-line drawings}\label{sse:straightline}

We first review results related to the problem of computing embedding-preserving straight-line drawings of $1$-plane graphs. More precisely, given a $1$-plane graph $G$, we say that $G$ has an \emph{embedding-preserving} straight-line drawing if there exists a ($1$-planar) straight-line drawing $\Gamma$ of $G$ that defines the same set of faces and the same outer face of $G$ (see also Section~\ref{se:preliminaries}).

\begin{figure}
\centering
\subfigure[]{\includegraphics[scale=1,page=1]{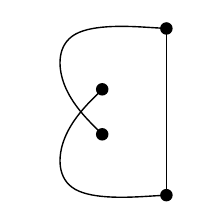}\label{fi:b-conf}}
\subfigure[]{\includegraphics[scale=1,page=2]{figs/badconf}\label{fi:w-conf}}
\subfigure[]{\includegraphics[scale=1,page=3]{figs/badconf}\label{fi:t-conf}}
\caption{(a) A B-configuration. (b) A W-configuration. (c) A T-configuration.\label{fi:bad-conf}}
\end{figure}

\begin{figure}
\centering
\subfigure[]{\includegraphics[scale=1,page=1]{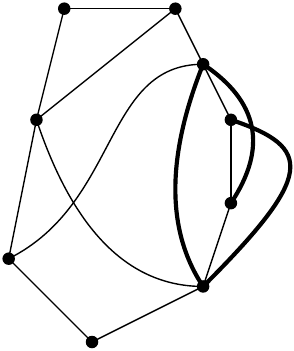}\label{fi:sl-drawing-1}}\hfil
\subfigure[]{\includegraphics[scale=1,page=2]{figs/sl-drawing}\label{fi:sl-drawing-2}}\hfil
\subfigure[]{\includegraphics[scale=1,page=3]{figs/sl-drawing}\label{fi:sl-drawing-3}}
\caption{(a) A $1$-plane graph containing a B-configuration (bold edges). (b) A different $1$-planar embedding of the graph in (a) where the B-configuration is removed. (c) A $1$-planar straight-line drawing that realized the embedding in (b).\label{fi:sl-drawing}}
\end{figure}

In 1988, Thomassen~\cite{Tho88} proved that a $1$-plane graph $G$ admits an embedding-preserving straight-line  drawing if and only if $G$ contains neither B-configurations nor W-configurations as subgraphs; see Figure~\ref{fi:bad-conf}. Based on this characterization, Hong \emph{et al.}~\cite{HELP12} described a linear-time algorithm to test whether a $1$-plane graph $G$ has an embedding-preserving straight-line drawing. The algorithm by Hong \emph{et al.}~\cite{HELP12} is based on an efficient procedure that checks whether $G$ contains any of the above mentioned forbidden configurations, and, if not, returns a valid drawing of $G$. The authors observed that the area requirement of embedding-preserving straight-line drawings of $1$-plane graphs may be exponential~\cite{HELP12}. More recently, Hong and Nagamochi~\cite{HongN16} proved that given a $1$-plane graph $G$, it can be tested in linear time whether a $1$-planar embedding of $G$ exists such that it preserves the pairs of crossing edges with respect to the original embedding and it can be realized as a straight-line drawing. Figure~\ref{fi:sl-drawing-1} shows a $1$-plane graph that does not admit an embedding-preserving straight-line drawing due to the presence of a B-configuration as a subgraph (drawn with bold edges). Figure~\ref{fi:sl-drawing-2} shows a different $1$-planar embedding of the same graph where the B-configuration is removed. Note that the graph in Figure~\ref{fi:sl-drawing-2} contains the same pairs of crossing edges as the graph in Figure~\ref{fi:sl-drawing-1}. Finally, Figure~\ref{fi:sl-drawing-3} shows a straight-line drawing that realizes the embedding in Figure~\ref{fi:sl-drawing-2}.

\begin{itemize}

\item \cite{HELP12}~S.~Hong, P.~Eades, G.~Liotta, and S.-H.~Poon.
\newblock F{\'a}ry's theorem for $1$-planar graphs.
\newblock In {\em {COCOON} 2012}, volume 7434 of {\em LNCS}, pages 335--346.  Springer, 2012.

\item \cite{HongN16}~S.~Hong and H.~Nagamochi.
\newblock Re-embedding a $1$-plane graph into a straight-line drawing in linear  time.
\newblock In {\em {GD} 2016}, volume 9801 of {\em LNCS}, pages 321--334.  Springer, 2016.

\item \cite{Tho88}~C.~Thomassen.
\newblock Rectilinear drawings of graphs.
\newblock {\em J.~Graph Theory}, 12(3):335--341, 1988.

\end{itemize}

In general, any $1$-planar graph admitting a $1$-planar straight-line drawing has at most $4n-9$ edges, and this bound is tight~\cite{D13}. Since $1$-planar graphs can have $4n-8$ edges, this implies that not all $1$-planar graphs admit a $1$-planar straight-line drawing, regardless of the embedding. Every $3$-connected $1$-planar graph $G$, however, does have a $1$-planar grid drawing such that all edges are drawn as straight-line segments, except for at most one edge on the outer face that requires one bend, as shown by Alam \emph{et al.}~\cite{ABK13}. The algorithm described in~\cite{ABK13} takes as input a $1$-plane graph but may produce a $1$-planar drawing that does not preserve the embedding of the input graph. This is due to a preprocessing step in which the graph is first augmented by adding edges such that the four end-vertices of each pair of crossing edges induce a $K_4$ (the missing edges can be added without introducing crossings in the drawing); and then any B-configuration is removed by rerouting one of its edges. Since $3$-connected $1$-planar graphs contain at most one W-configuration, the graph produced by this preprocessing step contains at most one forbidden configuration, i.e., at most one W-configuration on its outerface.  The area of the computed drawing is $O(n) \times O(n)$ and the algorithm runs in $O(n)$ time, where $n$ is the number of vertices of the input graph.

\begin{itemize}

\item \cite{ABK13}~M.~J.~Alam, F.~J.~Brandenburg, and S.~G.~Kobourov.
\newblock Straight-line grid drawings of 3-connected $1$-planar graphs.
\newblock In {\em {GD} 2013}, volume 8242 of {\em LNCS}, pages 83--94.  Springer, 2013.

\item \cite{D13}~W.~Didimo.
\newblock Density of straight-line $1$-planar graph drawings.
\newblock {\em Inf.~Process.~Lett.}, 113(7):236--240, 2013.

\end{itemize}

The triconnectivity requirement can be dropped if the input graph is IC-planar. More precisely, given an $n$-vertex IC-plane graph $G$, there exists an $O(n)$-time algorithm that computes  a $1$-planar straight-line drawing of $G$ on a grid of size $O(n^2) \times O(n^2)$~\cite{BrandenburgDEKL16}. The algorithm in~\cite{BrandenburgDEKL16} is based on an augmentation technique that makes $G$ $3$-connected by adding edges. This step might change the embedding of the input graph, but the new embedding is guaranteed to be IC-planar. Note that, unlike the case for general $1$-planar graphs, this implies that every IC-planar graph can be augmented to be $3$-connected without losing $1$-planarity.

\begin{itemize}

\item \cite{BrandenburgDEKL16}~F.~J.~Brandenburg, W.~Didimo, W.~S.~Evans, P.~Kindermann, G.~Liotta, and F.~Montecchiani.
\newblock Recognizing and drawing {IC}-planar graphs.
\newblock {\em Theor.~Comput.~Sci.}, 636:1--16, 2016.

\end{itemize}

Straight-line drawings of outer $1$-planar graphs have been studied by Di Giacomo \emph{et al.}~\cite{GiacomoLM15}. They proved that every outer $1$-planar graph $G$ with maximum vertex degree $\Delta$ admits an outer $1$-planar straight-line drawing $\Gamma$, with the additional property that $\Gamma$ uses $O(\Delta)$ different slopes for the edge segments. Also, drawing $\Gamma$ can be computed in $O(n)$ time, where $n$ is the number of vertices of $G$. Since outer $1$-planar graphs are planar graphs\footnote{More precisely, every outer $1$-planar graph can be drawn on the plane without crossings if we do not restrict the position of the vertices.}, Di Giacomo \emph{et al.}~\cite{GiacomoLM15} also proved that every outer $1$-planar graph $G$ with maximum vertex degree $\Delta$ admits a planar straight-line drawing that uses at most $O(\Delta^2)$ slopes. 

\begin{itemize}

\item \cite{GiacomoLM15}~E.~{Di Giacomo}, G.~Liotta, and F.~Montecchiani.
\newblock Drawing outer $1$-planar graphs with few slopes.
\newblock {\em J.~Graph Algorithms Appl.}, 19(2):707--741, 2015.

\end{itemize}

Erten and Kobourov~\cite{ErtenK05} showed that a 3-connected planar graph and its dual (which always form a $1$-planar graph) can be drawn on a grid of size $(2n - 2) \times (2n - 2)$, where $n$ is the total number of vertices in the graph and its dual. All the edges are drawn as straight-line segments except for one edge on the outer face, which is drawn using two segments. Also, each dual vertex lies inside its primal face, and a pair of edges cross if and only if the edges are a primal-dual pair. The algorithm runs in $O(n)$ time.

\begin{itemize}

\item~\cite{ErtenK05}~C.~Erten and S.~G. Kobourov.
\newblock Simultaneous embedding of a planar graph and its dual on the grid.
\newblock {\em Theory Comput. Syst.}, 38(3):313--327, 2005.

\end{itemize}

We conclude this section by observing that every $n$-vertex $1$-planar graph has a crossing-free  straight-line drawing in 3D with vertices placed at integer coordinates and such that the occupied volume is $O(n \log n)$. This result follows from the following argument. Every $n$-vertex $1$-planar graph has layered treewidth at most $12$~\cite{2015arXiv150604380D} and, as a consequence of a result by Dujmovi{\'c} \emph{et al.}~\cite{DujimovicMW2017}, it has track-number $O(\log n)$ (see~\cite{DujimovicMW2017} for the definition of track-number). This, together with the fact that every $1$-planar graph has a proper vertex coloring with at most six colors, imply the existence of the desired drawing, as proved by Dujmovi{\'c} and Wood~\cite{dujwood14}.

\begin{itemize}

\item \cite{2015arXiv150604380D}~V.~{Dujmovi{\'c}}, D.~{Eppstein}, and D.~R. {Wood}.
\newblock Structure of graphs with locally restricted crossings.
\newblock {\em SIAM J. Discrete Math.}, 31(2):805--824, 2017.

\item \cite{DujimovicMW2017}~V.~{Dujmovi{\'c}}, P.~{Morin}, and D.~R. {Wood}.
\newblock {Layered separators in minor-closed families with applications}.
\newblock {\em J. Combin. Theory Ser. B.}, 2017.
\newblock doi:10.1016/j.jctb.2017.05.006.

\item~\cite{dujwood14}~V.~Dujmovi{\'c} and D.~R. Wood.
\newblock Three-dimensional grid drawings with sub-quadratic volume.
\newblock In {\em Towards a Theory of Geometric Graphs}, volume 342 of {\em Contemporary Mathematics}, pages 55--66. AMS, 2014.

\end{itemize}

\subsection{RAC drawings}\label{sse:rac}

\begin{figure}
\centering
\includegraphics[scale=1]{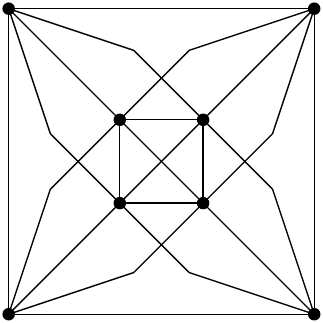}
\caption{A $1$-planar 1-bend RAC drawing.\label{fi:rac-example}}
\end{figure}

A \emph{$k$-bend Right-Angle-Crossing (RAC)  drawing} of a graph is a polyline drawing where each edge has at most $k$ bends and edges cross only at right angles. A $0$-bend RAC drawing is also called a \emph{straight-line RAC drawing}.  
Figure~\ref{fi:rac-example} shows a $1$-planar drawing in which each edge has at most one bend and crossings occur at right angles, i.e., a $1$-planar 1-bend RAC drawing. The study of RAC drawings is motivated by several cognitive experiments suggesting that while edge crossings in general make a drawing less readable, this effect is neutralized if edges cross at large angles~\cite{Huang07,HuangEH14,HuangHE08}.

\begin{itemize}

\item \cite{Huang07}~W.~Huang.
\newblock Using eye tracking to investigate graph layout effects.
\newblock In {\em {APVIS} 2007}, pages 97--100. IEEE, 2007.

\item \cite{HuangEH14}~W.~Huang, P.~Eades, and S.-H.~Hong.
\newblock Larger crossing angles make graphs easier to read.
\newblock {\em J.~Vis.~Lang.~Comput.}, 25(4):452--465, 2014.

\item \cite{HuangHE08}~W.~Huang, S.-H.~Hong, and P.~Eades.
\newblock Effects of crossing angles.
\newblock In {\em {PacificVis} 2008}, pages 41--46. IEEE, 2008.

\end{itemize}

\subsubsection{Straight-line RAC drawings}

It is known that any $n$-vertex graph that admits a straight-line RAC drawing has at most $4n-10$ edges and that this bound is tight~\cite{del-dgrac-2011}. This implies that there are $1$-planar graphs that admit a $1$-planar straight-line drawing but do not admit a straight-line RAC drawing. Moreover, it is known that graphs with a straight-line RAC drawing that are not $1$-planar do exist~\cite{el-rac1p-DAM13}. More recently, Bekos {\em et al.}~\cite{Bekos2017}  proved that deciding whether a graph has a $1$-planar straight-line RAC drawing is NP-hard in general and in the fixed-rotation-system setting.

\begin{itemize}

\item \cite{Bekos2017}~M.~A. Bekos, W.~Didimo, G.~Liotta, S.~Mehrabi, and F.~Montecchiani.
\newblock On {RAC} drawings of 1-planar graphs.
\newblock {\em Theor.~Comput.~Sci.}, 2017.
\newblock doi:10.1016/j.tcs.2017.05.039.

\item \cite{del-dgrac-2011}~W.~Didimo, P.~Eades, and G.~Liotta.
\newblock Drawing graphs with right angle crossings.
\newblock {\em Theor.~Comput.~Sci.}, 412(39):5156--5166, 2011.

\item \cite{el-rac1p-DAM13}~P.~Eades and G.~Liotta.
\newblock Right angle crossing graphs and $1$-planarity.
\newblock {\em Discrete Appl.~Math.}, 161(7-8):961--969, 2013.

\end{itemize}

The situation is different for IC-planar graphs and outer $1$-planar graphs. Brandenburg \emph{et al.}~\cite{BrandenburgDEKL16} proved that every $n$-vertex IC-planar graph has an IC-planar straight-line RAC drawing. The algorithm in~\cite{BrandenburgDEKL16} has $O(n^3)$ time complexity if an initial IC-planar embedding (which may be changed by the algorithm) is given as part of the input. The computed drawings may require exponential area and exponential area is sometimes necessary for IC-planar straight-line RAC drawings~\cite{BrandenburgDEKL16}. Finally, every outer $1$-planar graph $G$ has a straight-line RAC drawing that preserves the embedding of $G$~\cite{DehkordiE12}.

\begin{itemize}

\item \cite{BrandenburgDEKL16}~F.~J.~Brandenburg, W.~Didimo, W.~S.~Evans, P.~Kindermann, G.~Liotta, and  F.~Montecchiani.
\newblock Recognizing and drawing {IC}-planar graphs.
\newblock {\em Theor.~Comput.~Sci.}, 636:1--16, 2016.

\item \cite{DehkordiE12}~H.~R.~Dehkordi and P.~Eades.
\newblock Every outer-$1$-plane graph has a right angle crossing drawing.
\newblock {\em Int.~J.~Comput.~Geometry Appl.}, 22(6):543--558, 2012.

\end{itemize}

Brightwell and Scheinerman~\cite{BrightwellS93} proved that every 3-connected planar graph and its dual (which always form a $1$-planar graph) can be simultaneously drawn in the plane with straight-line edges so that the primal edges cross the dual edges at right
angles, provided that the vertex corresponding to the outer face is located at infinity. The result exploits the fact that every 3-connected planar graph $G$ can be represented as a collection of circles, one circle representing each vertex
and each face, so that, for each edge of $G$, the four circles representing the two endpoints and the two neighboring faces meet at a point. Moreover, the circles representing vertices cross the circles representing faces at right angles. Mohar~\cite{Mohar97} extends the results of Brightwell and Scheinerman by showing an approximation algorithm that given a 3-connected planar graph $G$ and a rational number $\varepsilon > 0$ finds an $\varepsilon$-approximation for the radii and the coordinates of the centers for the primal-dual circle representation for $G$, such that the angles of the primal-dual edge crossings are arbitrarily close to $\frac{\pi}{2}$. Neither of these two methods produce drawings in polynomial area.

\begin{itemize}

\item~\cite{BrightwellS93}~G.~R. Brightwell and E.~R. Scheinerman.
\newblock Representations of planar graphs.
\newblock {\em {SIAM} J. Discrete Math.}, 6(2):214--229, 1993.

\item~\cite{Mohar97}~B.~Mohar.
\newblock Circle packings of maps in polynomial time.
\newblock {\em Eur. J. Comb.}, 18(7):785--805, 1997.

\end{itemize}

\subsubsection{$k$-bend RAC drawings with $k>0$}

If we allow bends, it is known that every $1$-planar graph has a $1$-planar 1-bend RAC drawing~\cite{Bekos2017}. Specifically, there is an $O(n)$ time algorithm that takes as input an $n$-vertex $1$-plane graph $G$ and computes a $1$-planar 1-bend RAC drawing of $G$ (the  algorithm may change the $1$-planar embedding of $G$)~\cite{Bekos2017}. This algorithm may produce drawings that use exponential area. 

\begin{itemize}

\item \cite{Bekos2017}~M.~A. Bekos, W.~Didimo, G.~Liotta, S.~Mehrabi, and F.~Montecchiani.
\newblock On {RAC} drawings of 1-planar graphs.
\newblock {\em Theor.~Comput.~Sci.}, 2017.
\newblock doi:10.1016/j.tcs.2017.05.039.

\end{itemize}

Finally, Liotta and Montecchiani~\cite{LiottaM16} proved that every IC-plane graph has a $1$-planar 2-bend RAC drawing in quadratic area. Their technique is based on the construction of a visibility representation as intermediate step; see also Section~\ref{sse:visibility}.

\begin{itemize}

\item \cite{LiottaM16}~G.~Liotta and F.~Montecchiani.
\newblock {L}-visibility drawings of {IC}-planar graphs.
\newblock {\em Inf.~Process.~Lett.}, 116(3):217--222, 2016.

\end{itemize}

\subsection{Visibility representations}\label{sse:visibility}

A visibility representation of a graph $G$ maps the vertices of $G$ to geometric objects (such as bars or polygons), while the edges of $G$ are lines of sight between pairs of objects. The first visibility model studied in the literature are the bar visibility representations. In a \emph{bar visibility representation} of a graph $G$ each vertex $v$ of $G$ is mapped to a distinct horizontal segment $b(v)$ (called \emph{bar}) and each edge $(u,v)$ of $G$ corresponds to a vertical unobstructed segment (called \emph{visibility}) having one endpoint on $b(u)$ and the other one on $b(v)$. Such a representation is clearly planar (since edges are parallel segments) and every planar graph can be realized as a bar visibility representation~\cite{Duchet1983319,ov-grild-78,RosenstiehlT86,TamassiaTollis86,t-prg-84,Wismath85}.

\begin{itemize}

\item \cite{Duchet1983319}~P.~Duchet, Y.~O. Hamidoune, M.~L.~Vergnas, and H.~Meyniel.
\newblock Representing a planar graph by vertical lines joining different levels.
\newblock {\em Discrete Math.}, 46(3):319 -- 321, 1983.

\item \cite{ov-grild-78}~R.~H.~J.~M. Otten and J.~G.~V. Wijk.
\newblock Graph representations in interactive layout design.
\newblock In {\em {IEEE ISCSS}}, pages 914--918. IEEE, 1978.

\item \cite{RosenstiehlT86}~P.~Rosenstiehl and R.~E.~Tarjan.
\newblock Rectilinear planar layouts and bipolar orientations of planar graphs.
\newblock {\em Discr.~{\&} Comput.~Geom.}, 1:343--353, 1986.

\item \cite{TamassiaTollis86}~R.~Tamassia and I.~G.~Tollis.
\newblock A unified approach to visibility representations of planar graphs.
\newblock {\em Discr.~\& Comput.~Geom.}, 1(1):321--341, 1986.

\item \cite{t-prg-84}~C.~Thomassen.
\newblock Plane representations of graphs.
\newblock In {\em Progress in Graph Theory}, pages 43--69. AP, 1984.

\item \cite{Wismath85}~S.~K.~Wismath.
\newblock Characterizing bar line-of-sight graphs.
\newblock In {\em {SoCG} 1985}, pages 147--152. {ACM}, 1985.

\end{itemize}

In this section we consider four different visibility models used to represent $1$-planar graphs. The first model extends bar visibility representations by allowing edges to ``see'' through vertices. The other three visibility models extend bar visibility representations by allowing more complex shapes for the vertices and two orthogonal directions for the edges.

\subsubsection{Bar 1-visibility representations}
A \emph{bar $k$-visibility representation} is a bar visibility representation where each visibility can intersect at most $k$ bars~\cite{DeanEGLST07}. In other words, each edge can cross at most $k$ vertices. For values of $k$ greater than zero, this model allows for representing non-planar graphs. In particular, every $n$-vertex $1$-planar graph has a bar 1-visibility representation on a grid of size $O(n) \times O(n)$, as independently proved by Brandenburg~\cite{BrandenburgJGAA14} and by Evans \emph{et al.}~\cite{Evans0LMW14}. Both papers are based on constructive linear-time algorithms that take as input a $1$-plane graph $G$ and that may change the embedding of $G$ in order to construct the final representation.

\begin{itemize}

\item \cite{BrandenburgJGAA14}~F.~J.~Brandenburg.
\newblock 1-visibility representations of $1$-planar graphs.
\newblock {\em J.~Graph Algorithms Appl.}, 18(3):421--438, 2014.

\item \cite{DeanEGLST07}~A.~M.~Dean, W.~S.~Evans, E.~Gethner, J.~D.~Laison, M.~A.~Safari, and W.~T.~Trotter.
\newblock Bar $k$-visibility graphs.
\newblock {\em J.~Graph Algorithms Appl.}, 11(1):45--59, 2007.

\item \cite{Evans0LMW14}~W.~S.~Evans, M.~Kaufmann, W.~Lenhart, T.~Mchedlidze, and S.~K.~Wismath.
\newblock Bar 1-visibility graphs vs. other nearly planar graphs.
\newblock {\em J.~Graph Algorithms Appl.}, 18(5):721--739, 2014.

\end{itemize}

\subsubsection{L-visibility representations}

\newcommand{\SA}{\includegraphics{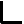}~}
\newcommand{\SB}{\includegraphics{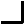}~}
\newcommand{\SC}{\includegraphics{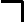}~}
\newcommand{\SD}{\includegraphics{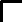}~}
\newcommand{\shapes}{$\{\SA,\SB,\SC,\SD\}$\xspace}

\begin{figure}
\centering
\subfigure[]{\includegraphics[scale=0.5,page=1]{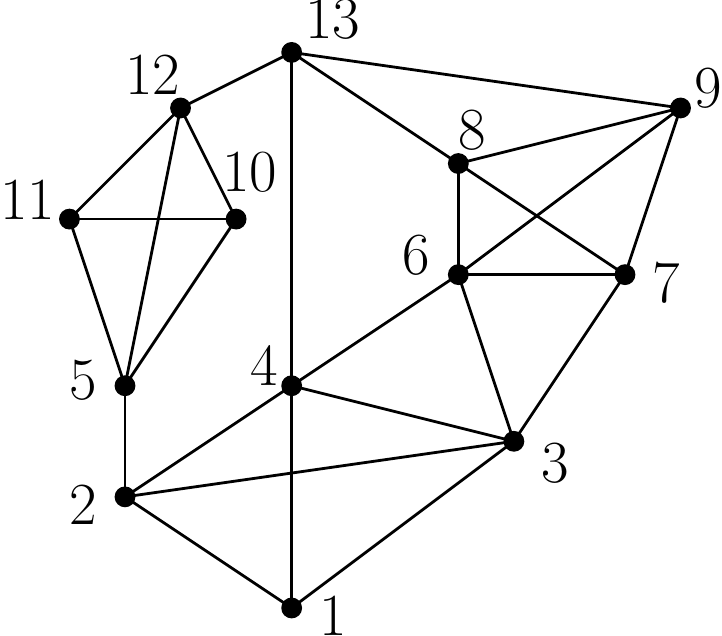}\label{fi:visrep-icpl-1}}\hfil
\subfigure[]{\includegraphics[scale=0.5,page=2]{figs/visrep-icplanar}\label{fi:visrep-icpl-2}}
\caption{(a) An IC-plane graph $G$ and (b) an L-visibility representation of $G$.\label{fi:visrep-icplanar}}
\end{figure}

In an {L-visibility representation} of a graph $G$, every vertex is represented by a horizontal and a vertical segment sharing an end-point (i.e., by an L-shape in the set \shapes), and each edge of $G$ is drawn as either a horizontal or a vertical visibility segment joining the two L-shapes corresponding to its two end-vertices. Either the vertical segment or the horizontal segment of an L-shape might have zero length. See Figure~\ref{fi:visrep-icplanar} for an illustration, and refer to:

\begin{itemize}

\item \cite{EvansLM16}~W.~S.~Evans, G.~Liotta, and F.~Montecchiani.
\newblock Simultaneous visibility representations of plane $st$-graphs using  {L}-shapes.
\newblock {\em Theor.~Comput.~Sci.}, 645:100--111, 2016.

\end{itemize}

Liotta and Montecchiani~\cite{LiottaM16} proved that every IC-plane graph $G$ with $n$ vertices admits an L-visibility representation in $O(n^2)$ area, which can be computed in $O(n)$ time. The algorithm may change the embedding of $G$, but the final representation is such that each visibility is crossed at most once and no two crossed visibilities are incident to the same L-shape.

\begin{itemize}

\item \cite{LiottaM16}~G.~Liotta and F.~Montecchiani.
\newblock {L}-visibility drawings of {IC}-planar graphs.
\newblock {\em Inf.~Process.~Lett.}, 116(3):217--222, 2016.

\end{itemize}

\subsubsection{Rectangle visibility representations}

\begin{figure}
\centering
\subfigure[]{\includegraphics[scale=1,page=1]{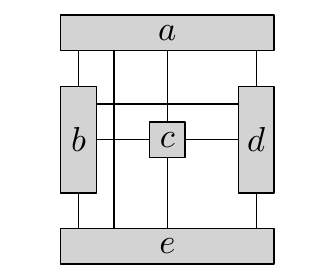}\label{fi:visrep-example-1}}\hfil
\subfigure[]{\includegraphics[scale=1,page=2]{figs/visrep-example}\label{fi:visrep-example-2}}\hfil
\subfigure[]{\includegraphics[scale=1,page=3]{figs/visrep-example}\label{fi:visrep-example-3}}
\caption{(a) A rectangle visibility representation of the complete graph $K_5$. A $1$-plane graph $G$ that does not admit a rectangle visibility representation due to the presence of a W-configuration (bold edges). (c) An ortho-polygon visibility representation of $G$ using at most one reflex corner per vertex.\label{fi:visrep-example}}
\end{figure}

A \emph{rectangle visibility representation} of a graph $G$ maps each vertex $v$ of $G$ to a distinct rectangle $r(v)$ and each edge $(u,v)$ of $G$ corresponds to a vertical or horizontal unobstructed segment (called \emph{visibility}) having one endpoint on $r(u)$ and the other one on $r(v)$. In this model horizontal and vertical visibilities may cross each other, whereas rectangles are not crossed. Given a rectangle visibility representation, we can extract a  drawing from it as follows. For each vertex $v$, place a point inside the corresponding rectangle $r(v)$ and connect it to all the attachment points of the visibilities on the boundary of $r(v)$; this can be done without creating any crossing and preserving the circular order of the edges around the vertices.  An embedded graph $G$ has an \emph{embedding-preserving} rectangle visibility representation $\Gamma$ if the drawing extracted from $\Gamma$ preserves the embedding of $G$.

Recently, Biedl \emph{et al.}~\cite{BiedlLM16} proved that a $1$-plane graph $G$ with $n$ vertices admits an embedding-preserving rectangle visibility representation  if and only if $G$ does not contain B-configurations, W-configurations,  and T-configurations; see Figure~\ref{fi:bad-conf}. 
The absence of these three configurations can be checked in $O(n)$ time, and if $G$ contains none of them, then an embedding-preserving rectangle visibility representation of $G$ can be computed in $O(n)$ time on a grid of size $O(n) \times O(n)$~\cite{BiedlLM16}. For example, Figure~\ref{fi:visrep-example-1} shows a rectangle visibility representation of the complete graph $K_5$, while Figure~\ref{fi:visrep-example-2} shows a $1$-plane graph that does not admit an embedding-preserving rectangle visibility representation due to the presence of a W-configuration as a subgraph. Moreover, there exist $1$-planar graphs that do not admit rectangle visibility representations (for any $1$-planar embedding), unless a linear number of edges is removed~\cite{BiedlLM16}.

\begin{itemize}

\item \cite{BiedlLM16}~T.~C.~Biedl, G.~Liotta, and F.~Montecchiani.
\newblock On visibility representations of non-planar graphs.
\newblock In {\em {SoCG} 2016}, volume~51 of {\em LIPIcs}, pages 19:1--19:16.  Schloss Dagstuhl - Leibniz-Zentrum fuer Informatik, 2016.

\end{itemize}

In general, the complexity of testing whether a $1$-planar graph admits a rectangle visibility representation is unknown, although the problem is NP-hard in general~\cite{Shermer96}, and it likely to remain NP-hard even for $1$-planar graphs.

\begin{itemize}

\item \cite{Shermer96}~T.~C.~Shermer.
\newblock On rectangle visibility graphs. {III.} {E}xternal visibility and  complexity.
\newblock In {\em {CCCG} 1996}, pages 234--239. Carleton University Press,  1996.

\end{itemize}

\subsubsection{Ortho-polygon visibility representations}

Di Giacomo \emph{et al.}~\cite{DiGiacomo2017} introduced \emph{ortho-polygon visibility representations (OPVR)}, where vertices can be orthogonal polygons rather than just rectangles, while edges are still horizontal or vertical segments between the corresponding pairs of polygons.

It is proved that every $1$-plane graph $G$ with $n$ vertices admits an embedding-preserving OPVR on a grid of size $O(n) \times O(n)$, which can be computed in $O(n)$ time~\cite{DiGiacomo2017}. (The notion of embedding-preserving OPVR  can be defined similarly as for rectangle visibility representations.) For example, Figure~\ref{fi:visrep-example-2} shows a $1$-plane graph (that does not admit an embedding-preserving rectangle visibility representation), and Figure~\ref{fi:visrep-example-3} shows an OPVR of $G$. In order to control the complexity of the polygons representing the vertices in an OPVR, Di Giacomo \emph{et al.}~\cite{DiGiacomo2017} proved that if $G$ is $3$-connected, then $G$ admits an embedding-preserving OPVR with at most 12 reflex corners per vertex. The proof is based on an algorithm that exploits the results on the edge partitions of $3$-connected $1$-plane graphs described in Section~\ref{sse:edgepartitions}. Also, there exist $3$-connected $1$-plane graphs that require at least 2 reflex corners on some vertices in any embedding-preserving OPVR. Moreover, there are 2-connected $1$-planar graphs that require linearly many reflex corners on some vertices in any OPVR (regardless of the $1$-planar embedding)~\cite{DiGiacomo2017}. In general, given an $n$-vertex $1$-plane graph $G$, computing an embedding-preserving OPVR of $G$ with the minimum number of reflex corners per vertex can be done in $O(n^\frac{7}{4}\sqrt{\log n})$ time~\cite{DiGiacomo2017}.

\begin{itemize}

\item \cite{DiGiacomo2017}~E.~{Di Giacomo}, W.~Didimo, W.~S. Evans, G.~Liotta, H.~Meijer, F.~Montecchiani, and S.~K. Wismath.
\newblock Ortho-polygon visibility representations of embedded graphs.
\newblock {\em Algorithmica}, 2017.

\end{itemize}

\subsection{Contact representations}\label{sse:contact}

Alam \emph{et al.}~\cite{AlamEKPTU15} studied \emph{contact representations} of graphs in which vertices are represented by axis-aligned polyhedra in 3D and edges are realized by non-zero area common boundaries between corresponding polyhedra. In particular, a \emph{box-contact representation} are contact representations where vertices are axis-aligned boxes. The authors described an $O(n)$-time algorithm for computing box-contact representations of $n$-vertex optimal $1$-planar graphs without separating 4-cycles. Since not every optimal $1$-planar graph admits a box-contact representation, the authors also considered contact representations where vertices are L-shaped polyhedra, called \emph{L-contact representations}, and they provided a $O(n^2)$-time algorithm for representing every $n$-vertex optimal $1$-planar graph with this model.

\begin{itemize}

\item \cite{AlamEKPTU15}~M.~J.~Alam, W.~S.~Evans, S.~G.~Kobourov, S.~Pupyrev, J.~Toeniskoetter, and  T.~Ueckerdt.
\newblock Contact representations of graphs in {3D}.
\newblock In {\em {WADS} 2015}, volume 9214 of {\em LNCS}, pages 14--27.  Springer, 2015.

\end{itemize}

\section{Final Remarks and Open Problems}\label{se:remarks}
In this paper we presented an annotated bibliography that surveys the state of the art of the combinatorial, geometric and algorithmic study of $1$-planar graphs. We divided the covered results in three main sections, namely, results related with the problems of recognizing and characterizing $1$-planar graphs, results that investigate structural properties of $1$-planar graphs, and results that study geometric representations of $1$-planar graphs.

The family if $1$-planar graphs fits in the more general framework of ``beyond planarity'', in which families of non-planar graphs are defined based on forbidden edge crossing configurations. For further details on this topic we refer the reader to the paper by Liotta~\cite{Liotta14}:

\begin{itemize}

\item \cite{Liotta14}
G.~Liotta.
\newblock Graph drawing beyond planarity: some results and open problems.
\newblock In {\em {ICTCS} 2014.}, volume 1231 of {\em {CEUR} Workshop  Proceedings}, pages 3--8. CEUR-WS.org, 2014.

\end{itemize}

We conclude with a selection of open problems, grouped according to the main sections of this paper.

\smallskip

Open problems related with the topics covered in Section~\ref{se:characterization-rec}.
		\begin{enumerate}
			\item What classes of $1$-planar graphs are recognizable in polynomial time with or without a rotation system? In particular, what is the complexity of recognizing maximal $1$-planar graphs without a fixed rotation system?
		
			\item What is the complexity of recognizing $1$-planar graphs with bounded vertex degree $\Delta$, say with $\Delta=3$?
			
			\item It would be interesting to design and evaluate practical algorithms for recognizing (classes of) $1$-planar graphs. 

		\end{enumerate}
		
Open problems related with the topics covered in Section~\ref{se:properties}.
		\begin{enumerate}
		
		\item Is it possible to compute a proper vertex coloring with at most 6 colors of any $n$-vertex optimal $1$-planar graph in $O(n)$ time?

            \item What is the complexity of deciding whether a $1$-planar graph has a proper vertex coloring with at most 5 colors?

			\item Every $1$-planar graph admits an edge partition such that the red graph is a forest. However, nothing can be said on the structure induced by those crossing edges colored blue. Can we color with two colors the crossing edges of a $1$-plane graph such that each monochromatic set induces a forest?
			
			\item The \emph{star arboricity} of a graph is the minimum number of edge-disjoint forests that cover the graph, such that each tree of each forest is a star. Planar graphs have star arboricity 5~\cite{HAKIMI199693}. What is the star arboricity of $1$-planar graphs? Note that it must be between 5~and~7.
			
			\item An $n$-vertex planar graph $G$ with minimum vertex degree 3 contains a matching of size at least $(n+2)/3$~\cite{NishizekiB79}. Can we establish a similar lower bound for the cardinality of maximum matchings in $1$-planar graphs?
			
			\item What is the book thickness of (optimal) $1$-planar graphs?

		\end{enumerate}

Open problems related with the topics covered in Section~\ref{se:geometric-representations}.
		\begin{enumerate}
			\item Can we characterize and/or recognize those $1$-planar graphs that admit a straight-line drawing with (strictly) convex faces?
			
			\item Is there a value of $\Delta$ such that every $1$-planar graph with maximum vertex degree $\Delta$ admits a $1$-planar straight-line RAC drawing?
	
			\item The family of graphs that admit a bar 1-visibility representation includes $1$-planar graphs. 	What is the complexity of recognizing graphs in this family?
			
			\item Every 3-connected $1$-plane graph admits an embedding-preserving ortho-polygon visibility representation with at most 12 reflex corners, while 2 reflex corners are sometimes needed on some vertices. 
			Is 2 a tight bound?
			
			\item What classes of $1$-planar graphs, other than the optimal ones, admit an L-contact representation?
			
			\item Apart from the results in Sections~\ref{sse:straightline} and~\ref{sse:contact}, very little is known in terms of 3-dimensional representations of $1$-planar graphs. We believe that this topic deserves further attention.
			
		\end{enumerate}

\section*{Acknowledgments}
We thank Jawaherul Alam, Michael A. Bekos, Franz Brandenburg, Walter Didimo, David Eppstein, Seok-Hee Hong, Michael Kaufmann, Sergey Pupyrev, and David R. Wood for great discussions, explanations, and references.

{\small \bibliography{references}}

\begin{thebibliography}{100}

\bibitem{Ackerman14}
E.~Ackerman.
\newblock A note on 1-planar graphs.
\newblock {\em Discrete Appl. Math.}, 175:104--108, 2014.

\bibitem{ABK13}
M.~J. Alam, F.~J. Brandenburg, and S.~G. Kobourov.
\newblock Straight-line grid drawings of 3-connected 1-planar graphs.
\newblock In {\em {GD} 2013}, volume 8242 of {\em LNCS}, pages 83--94.
  Springer, 2013.

\bibitem{AlamBK15}
M.~J. Alam, F.~J. Brandenburg, and S.~G. Kobourov.
\newblock On the book thickness of 1-planar graphs.
\newblock {\em CoRR}, abs/1510.05891, 2015.

\bibitem{AlamEKPTU15}
M.~J. Alam, W.~S. Evans, S.~G. Kobourov, S.~Pupyrev, J.~Toeniskoetter, and
  T.~Ueckerdt.
\newblock Contact representations of graphs in {3D}.
\newblock In {\em {WADS} 2015}, volume 9214 of {\em LNCS}, pages 14--27.
  Springer, 2015.

\bibitem{AMC10}
M.~O. Albertson.
\newblock Chromatic number, independence ratio, and crossing number.
\newblock {\em Ars Math. Contemp.}, 1(1), 2008.

\bibitem{Angelini2012}
P.~Angelini, G.~{Di Battista}, F.~Frati, M.~Patrignani, and I.~Rutter.
\newblock Testing the simultaneous embeddability of two graphs whose
  intersection is a biconnected or a connected graph.
\newblock {\em J. Discrete Algorithms}, 14:150 -- 172, 2012.

\bibitem{MR1025335}
K.~Appel and W.~Haken.
\newblock {\em Every planar map is four colorable}, volume~98 of {\em
  Contemporary Mathematics}.
\newblock AMS, 1989.

\bibitem{AuerBBGHNR16}
C.~Auer, C.~Bachmaier, F.~J. Brandenburg, A.~Glei{\ss}ner, K.~Hanauer,
  D.~Neuwirth, and J.~Reislhuber.
\newblock Outer 1-planar graphs.
\newblock {\em Algorithmica}, 74(4):1293--1320, 2016.

\bibitem{AuerBGR15}
C.~Auer, F.~J. Brandenburg, A.~Glei{\ss}ner, and J.~Reislhuber.
\newblock 1-planarity of graphs with a rotation system.
\newblock {\em J. Graph Algorithms Appl.}, 19(1):67--86, 2015.

\bibitem{BachmaierBHNR17}
C.~Bachmaier, F.~J. Brandenburg, K.~Hanauer, D.~Neuwirth, and J.~Reislhuber.
\newblock Nic-planar graphs.
\newblock {\em CoRR}, abs/1701.04375, 2017.

\bibitem{Baker1994}
B.~S. Baker.
\newblock Approximation algorithms for {NP}-complete problems on planar graphs.
\newblock {\em J. ACM}, 41(1):153--180, 1994.

\bibitem{BCE13}
M.~J. Bannister, S.~Cabello, and D.~Eppstein.
\newblock Parameterized complexity of 1-planarity.
\newblock In {\em {WADS} 2013}, volume 8037 of {\em LNCS}, pages 97--108.
  Springer, 2013.

\bibitem{Baur2005}
M.~Baur and U.~Brandes.
\newblock Crossing reduction in circular layouts.
\newblock In {\em {WG} 2004}, volume 3353 of {\em LNCS}, pages 332--343.
  Springer, 2005.

\bibitem{Behzad65}
M.~Behzad.
\newblock {\em Graphs and Their Chromatic Numbers}.
\newblock PhD thesis, Michigan State University, Department of Mathematics,
  1965.

\bibitem{BekosB0R15}
M.~A. Bekos, T.~Bruckdorfer, M.~Kaufmann, and C.~N. Raftopoulou.
\newblock 1-planar graphs have constant book thickness.
\newblock In {\em {ESA} 2015}, volume 9294 of {\em LNCS}, pages 130--141.
  Springer, 2015.

\bibitem{Bekos2017}
M.~A. Bekos, W.~Didimo, G.~Liotta, S.~Mehrabi, and F.~Montecchiani.
\newblock On {RAC} drawings of 1-planar graphs.
\newblock {\em Theor. Comput. Sci.}, 2017.
\newblock doi:10.1016/j.tcs.2017.05.039.

\bibitem{Bekos0Z15}
M.~A. Bekos, M.~Kaufmann, and C.~Zielke.
\newblock The book embedding problem from a {SAT}-solving perspective.
\newblock In {\em {GD} 2015}, volume 9411 of {\em LNCS}, pages 125--138.
  Springer, 2015.

\bibitem{BiedlLM16}
T.~C. Biedl, G.~Liotta, and F.~Montecchiani.
\newblock On visibility representations of non-planar graphs.
\newblock In {\em {SoCG} 2016}, volume~51 of {\em LIPIcs}, pages 19:1--19:16.
  Schloss Dagstuhl - Leibniz-Zentrum fuer Informatik, 2016.

\bibitem{BSW83}
R.~Bodendiek, H.~Schumacher, and K.~Wagner.
\newblock Bemerkungen zu einem {S}echsfarbenproblem von {G.} {R}ingel.
\newblock {\em Abhandlungen aus dem Mathematischen Seminar der Universitaet
  Hamburg}, 53(1):41--52, 1983.

\bibitem{BSW84}
R.~Bodendiek, H.~Schumacher, and K.~Wagner.
\newblock Uber 1-optimale {G}raphen.
\newblock {\em Mathematische Nachrichten}, 117(1):323--339, 1984.

\bibitem{bm-gt-07}
J.~A. Bondy and U.~S.~R. Murty.
\newblock {\em Graph theory}.
\newblock Graduate texts in mathematics. Springer, 2007.

\bibitem{BKRS01}
O.~Borodin, A.~Kostochka, A.~Raspaud, and E.~Sopena.
\newblock Acyclic colouring of 1-planar graphs.
\newblock {\em Discrete Appl. Math.}, 114(1–3):29 -- 41, 2001.

\bibitem{B84}
O.~V. Borodin.
\newblock Solution of the ringel problem on vertex-face coloring of planar
  graphs and coloring of $1$-planar graphs.
\newblock {\em Metody Diskret. Analiz}, 108:12--26, 1984.

\bibitem{B95}
O.~V. Borodin.
\newblock A new proof of the 6 color theorem.
\newblock {\em J. Graph Theory}, 19(4):507--521, 1995.

\bibitem{BrandenburgJGAA14}
F.~J. Brandenburg.
\newblock 1-visibility representations of 1-planar graphs.
\newblock {\em J. Graph Algorithms Appl.}, 18(3):421--438, 2014.

\bibitem{Brandenburg15}
F.~J. Brandenburg.
\newblock On 4-map graphs and 1-planar graphs and their recognition problem.
\newblock {\em CoRR}, abs/1509.03447, 2015.

\bibitem{Brandenburg16b}
F.~J. Brandenburg.
\newblock Recognizing {IC}-planar and {NIC}-planar graphs.
\newblock {\em CoRR}, abs/1610.08884, 2016.

\bibitem{Brandenburg16a}
F.~J. Brandenburg.
\newblock Recognizing optimal 1-planar graphs in linear time.
\newblock {\em Algorithmica}, 2016.
\newblock doi:10.1007/s00453-016-0226-8.

\bibitem{BrandenburgDEKL16}
F.~J. Brandenburg, W.~Didimo, W.~S. Evans, P.~Kindermann, G.~Liotta, and
  F.~Montecchiani.
\newblock Recognizing and drawing {IC}-planar graphs.
\newblock {\em Theor. Comput. Sci.}, 636:1--16, 2016.

\bibitem{BEGGHR13}
F.-J. Brandenburg, D.~Eppstein, A.~Glei{\ss}ner, M.~T. Goodrich, K.~Hanauer,
  and J.~Reislhuber.
\newblock On the density of maximal 1-planar graphs.
\newblock In {\em {GD} 2012}, volume 7704 of {\em LNCS}, pages 327--338.
  Springer, 2013.

\bibitem{BrightwellS93}
G.~R. Brightwell and E.~R. Scheinerman.
\newblock Representations of planar graphs.
\newblock {\em {SIAM} J. Discrete Math.}, 6(2):214--229, 1993.

\bibitem{BGGMTW05}
G.~Brinkmann, S.~Greenberg, C.~S. Greenhill, B.~D. McKay, R.~Thomas, and
  P.~Wollan.
\newblock Generation of simple quadrangulations of the sphere.
\newblock {\em Discrete Math.}, 305(1--3):33--54, 2005.

\bibitem{Bucko2015}
J.~Bucko and J.~Czap.
\newblock 1-planar lexicographic products of graphs.
\newblock {\em Applied Mathematical Sciences}, 9(109):5441--5449, 2015.

\bibitem{CM12}
S.~Cabello and B.~Mohar.
\newblock Adding one edge to planar graphs makes crossing number and
  1-planarity hard.
\newblock {\em {SIAM} J. Comput.}, 42(5):1803--1829, 2013.

\bibitem{Chaitin1982}
G.~J. Chaitin.
\newblock Register allocation \& spilling via graph coloring.
\newblock {\em SIGPLAN Not.}, 17(6):98--101, 1982.

\bibitem{Chang2013}
G.~J. Chang.
\newblock {\em Algorithmic Aspects of Domination in Graphs}, pages 221--282.
\newblock Springer, New York, NY, 2013.

\bibitem{ChenGP98}
Z.~Chen, M.~Grigni, and C.~H. Papadimitriou.
\newblock Planar map graphs.
\newblock In {\em {STOC} 1998}, pages 514--523. {ACM}, 1998.

\bibitem{ChenGP02}
Z.~Chen, M.~Grigni, and C.~H. Papadimitriou.
\newblock Map graphs.
\newblock {\em J. {ACM}}, 49(2):127--138, 2002.

\bibitem{ChenGP06}
Z.~Chen, M.~Grigni, and C.~H. Papadimitriou.
\newblock Recognizing hole-free 4-map graphs in cubic time.
\newblock {\em Algorithmica}, 45(2):227--262, 2006.

\bibitem{ChenK05}
Z.~Chen and M.~Kouno.
\newblock A linear-time algorithm for 7-coloring 1-plane graphs.
\newblock {\em Algorithmica}, 43(3):147--177, 2005.

\bibitem{ChenZ07}
Z.-Z. Chen.
\newblock New bounds on the edge number of a $k$-map graph.
\newblock {\em J. Graph Theory}, 55(4):267--290, 2007.

\bibitem{Chepoi2002}
V.~Chepoi, F.~Dragan, and Y.~Vax\`{e}s.
\newblock Center and diameter problems in plane triangulations and
  quadrangulations.
\newblock In {\em {ACM-SIAM SODA 2002}}, pages 346--355. SIAM, 2002.

\bibitem{ChungLR1987}
F.~R.~K. Chung, F.~T. Leighton, and A.~L. Rosenberg.
\newblock Embedding graphs in books: A layout problem with applications to
  {VLSI} design.
\newblock {\em {SIAM} J. Algebraic Discr. Meth.}, 8(1):33--58, 1987.

\bibitem{C13}
J.~Czap.
\newblock A note on total colorings of 1-planar graphs.
\newblock {\em Inf. Process. Lett.}, 113(14-16):516--517, 2013.

\bibitem{CzapH12}
J.~Czap and D.~Hud{\'{a}}k.
\newblock 1-planarity of complete multipartite graphs.
\newblock {\em Discrete Appl. Math.}, 160(4-5):505--512, 2012.

\bibitem{CH13}
J.~Czap and D.~Hud{\'a}k.
\newblock On drawings and decompositions of 1-planar graphs.
\newblock {\em Electr. J. Comb.}, 20(2):P54, 2013.

\bibitem{Czap2014}
J.~Czap, D.~Hud{\'a}k, and T.~Madaras.
\newblock Joins of 1-planar graphs.
\newblock {\em Acta Math. Sin. English Series}, 30(11):1867--1876, 2014.

\bibitem{CzapPS16}
J.~Czap, J.~Przybylo, and E.~Skrabul'{\'{a}}kov{\'{a}}.
\newblock On an extremal problem in the class of bipartite 1-planar graphs.
\newblock {\em Discuss. Math. Graph Theory}, 36(1):141--151, 2016.

\bibitem{CzapS17}
J.~{Czap} and P.~{{\v S}ugerek}.
\newblock Drawing graph joins in the plane with restrictions on crossings.
\newblock {\em Filomat}, 31(2):363–--370, 2017.

\bibitem{DeanEGLST07}
A.~M. Dean, W.~S. Evans, E.~Gethner, J.~D. Laison, M.~A. Safari, and W.~T.
  Trotter.
\newblock Bar $k$-visibility graphs.
\newblock {\em J. Graph Algorithms Appl.}, 11(1):45--59, 2007.

\bibitem{DehkordiE12}
H.~R. Dehkordi and P.~Eades.
\newblock Every outer-1-plane graph has a right angle crossing drawing.
\newblock {\em Int. J. Comput. Geometry Appl.}, 22(6):543--558, 2012.

\bibitem{BattistaETT94}
G.~{Di Battista}, P.~Eades, R.~Tamassia, and I.~G. Tollis.
\newblock Algorithms for drawing graphs: an annotated bibliography.
\newblock {\em Comput. Geom.}, 4:235--282, 1994.

\bibitem{dett-gdavg-99}
G.~{Di Battista}, P.~Eades, R.~Tamassia, and I.~G. Tollis.
\newblock {\em Graph Drawing: Algorithms for the Visualization of Graphs}.
\newblock Prentice-Hall, 1999.

\bibitem{DiGiacomoEtAl2017}
E.~{Di Giacomo}, W.~Didimo, W.~S. Evans, G.~Liotta, H.~Meijer, F.~Montecchiani,
  and S.~K. Wismath.
\newblock New results on edge partitions of 1-plane graphs.
\newblock {\em CoRR}, abs/1706.05161, 2017.

\bibitem{DiGiacomo2017}
E.~{Di Giacomo}, W.~Didimo, W.~S. Evans, G.~Liotta, H.~Meijer, F.~Montecchiani,
  and S.~K. Wismath.
\newblock Ortho-polygon visibility representations of embedded graphs.
\newblock {\em Algorithmica}, 2017.
\newblock doi:10.1007/s00453-017-0324-2.

\bibitem{GiacomoLM15}
E.~{Di Giacomo}, G.~Liotta, and F.~Montecchiani.
\newblock Drawing outer 1-planar graphs with few slopes.
\newblock {\em J. Graph Algorithms Appl.}, 19(2):707--741, 2015.

\bibitem{D13}
W.~Didimo.
\newblock Density of straight-line 1-planar graph drawings.
\newblock {\em Inf. Process. Lett.}, 113(7):236--240, 2013.

\bibitem{del-dgrac-2011}
W.~Didimo, P.~Eades, and G.~Liotta.
\newblock Drawing graphs with right angle crossings.
\newblock {\em Theor. Comput. Sci.}, 412(39):5156--5166, 2011.

\bibitem{Diestel12}
R.~Diestel.
\newblock {\em Graph Theory, 4th Edition}, volume 173 of {\em Graduate texts in
  mathematics}.
\newblock Springer, 2012.

\bibitem{Ding2000221}
G.~Ding, B.~Oporowski, D.~P. Sanders, and D.~Vertigan.
\newblock Surfaces, tree-width, clique-minors, and partitions.
\newblock {\em J. Combin. Theory Ser. B}, 79(2):221 -- 246, 2000.

\bibitem{DowneyF99}
R.~G. Downey and M.~R. Fellows.
\newblock {\em Parameterized Complexity}.
\newblock Monographs in Computer Science. Springer, 1999.

\bibitem{Duchet1983319}
P.~Duchet, Y.~O. Hamidoune, M.~L. Vergnas, and H.~Meyniel.
\newblock Representing a planar graph by vertical lines joining different
  levels.
\newblock {\em Discrete Math.}, 46(3):319 -- 321, 1983.

\bibitem{2015arXiv150604380D}
V.~Dujmovi{\'c}, D.~Eppstein, and D.~R. Wood.
\newblock Structure of graphs with locally restricted crossings.
\newblock {\em SIAM J. Discrete Math.}, 31(2):805--824, 2017.

\bibitem{DujimovicMW2017}
V.~{Dujmovi{\'c}}, P.~{Morin}, and D.~{Wood}.
\newblock {Layered separators in minor-closed families with applications}.
\newblock {\em J. Combin. Theory Ser. B.}, 2017.
\newblock doi:10.1016/j.jctb.2017.05.006.

\bibitem{dujwood14}
V.~Dujmovi{\'c} and D.~R. Wood.
\newblock Three-dimensional grid drawings with sub-quadratic volume.
\newblock In {\em Towards a Theory of Geometric Graphs}, volume 342 of {\em
  Contemporary Mathematics}, pages 55--66. AMS, 2014.

\bibitem{EHKLSS13}
P.~Eades, S.~Hong, N.~Katoh, G.~Liotta, P.~Schweitzer, and Y.~Suzuki.
\newblock Testing maximal 1-planarity of graphs with a rotation system in
  linear time.
\newblock In {\em {GD} 2012}, volume 7704 of {\em LNCS}, pages 339--345.
  Springer, 2013.

\bibitem{el-rac1p-DAM13}
P.~Eades and G.~Liotta.
\newblock Right angle crossing graphs and 1-planarity.
\newblock {\em Discrete Appl. Math.}, 161(7-8):961--969, 2013.

\bibitem{ec-pepg+-88}
E.~S. Elmallah and C.~J. Colbourn.
\newblock Partitioning the edges of a planar graph into two partial $k$-trees.
\newblock In {\em {CGTC} 1988}, Congressus numerantium, pages 69--80. Utilitas
  Mathematica, 1988.

\bibitem{ErtenK05}
C.~Erten and S.~G. Kobourov.
\newblock Simultaneous embedding of a planar graph and its dual on the grid.
\newblock {\em Theory Comput. Syst.}, 38(3):313--327, 2005.

\bibitem{Evans0LMW14}
W.~S. Evans, M.~Kaufmann, W.~Lenhart, T.~Mchedlidze, and S.~K. Wismath.
\newblock Bar 1-visibility graphs vs. other nearly planar graphs.
\newblock {\em J. Graph Algorithms Appl.}, 18(5):721--739, 2014.

\bibitem{EvansLM16}
W.~S. Evans, G.~Liotta, and F.~Montecchiani.
\newblock Simultaneous visibility representations of plane $st$-graphs using
  {L}-shapes.
\newblock {\em Theor. Comput. Sci.}, 645:100--111, 2016.

\bibitem{FM07}
I.~Fabrici and T.~Madaras.
\newblock The structure of 1-planar graphs.
\newblock {\em Discrete Math.}, 307(7--8):854--865, 2007.

\bibitem{FlumG06}
J.~Flum and M.~Grohe.
\newblock {\em Parameterized Complexity Theory}.
\newblock Texts in Theoretical Computer Science. An {EATCS} Series. Springer,
  2006.

\bibitem{FoxPS13}
J.~Fox, J.~Pach, and A.~Suk.
\newblock The number of edges in $k$-quasi-planar graphs.
\newblock {\em {SIAM} J. Discrete Math.}, 27(1):550--561, 2013.

\bibitem{GareyJ79}
M.~R. Garey and D.~S. Johnson.
\newblock {\em Computers and Intractability: {A} Guide to the Theory of
  {NP}-Completeness}.
\newblock W. H. Freeman, 1979.

\bibitem{g-agt-85}
A.~Gibbons.
\newblock {\em {Algorithmic Graph Theory}}.
\newblock Cambridge University Press, 1985.

\bibitem{Goncalves05}
D.~Gon{\c{c}}alves.
\newblock Edge partition of planar graphs into two outerplanar graphs.
\newblock In {\em {STOC} 2005}, pages 504--512. {ACM}, 2005.

\bibitem{GB07}
A.~Grigoriev and H.~L. Bodlaender.
\newblock Algorithms for graphs embeddable with few crossings per edge.
\newblock {\em Algorithmica}, 49(1):1--11, 2007.

\bibitem{HAKIMI199693}
S.~Hakimi, J.~Mitchem, and E.~Schmeichel.
\newblock Star arboricity of graphs.
\newblock {\em Discrete Math.}, 149(1):93 -- 98, 1996.

\bibitem{Har-PeledQ15}
S.~Har{-}Peled and K.~Quanrud.
\newblock Approximation algorithms for polynomial-expansion and low-density
  graphs.
\newblock In {\em {ESA} 2015}, volume 9294 of {\em LNCS}, pages 717--728.
  Springer, 2015.

\bibitem{h-gt-72}
F.~Harary, editor.
\newblock {\em Graph Theory}.
\newblock Addison-Wesley, 1972.

\bibitem{HongEKLSS15}
S.~Hong, P.~Eades, N.~Katoh, G.~Liotta, P.~Schweitzer, and Y.~Suzuki.
\newblock A linear-time algorithm for testing outer-1-planarity.
\newblock {\em Algorithmica}, 72(4):1033--1054, 2015.

\bibitem{HELP12}
S.~Hong, P.~Eades, G.~Liotta, and S.-H. Poon.
\newblock F{\'a}ry's theorem for 1-planar graphs.
\newblock In {\em {COCOON} 2012}, volume 7434 of {\em LNCS}, pages 335--346.
  Springer, 2012.

\bibitem{HongN16}
S.~Hong and H.~Nagamochi.
\newblock Re-embedding a 1-plane graph into a straight-line drawing in linear
  time.
\newblock In {\em {GD} 2016}, volume 9801 of {\em LNCS}, pages 321--334.
  Springer, 2016.

\bibitem{Huang07}
W.~Huang.
\newblock Using eye tracking to investigate graph layout effects.
\newblock In {\em {APVIS} 2007}, pages 97--100. IEEE, 2007.

\bibitem{HuangEH14}
W.~Huang, P.~Eades, and S.~Hong.
\newblock Larger crossing angles make graphs easier to read.
\newblock {\em J. Vis. Lang. Comput.}, 25(4):452--465, 2014.

\bibitem{HuangHE08}
W.~Huang, S.~Hong, and P.~Eades.
\newblock Effects of crossing angles.
\newblock In {\em {PacificVis} 2008}, pages 41--46. IEEE, 2008.

\bibitem{HM09}
D.~Hud{\'a}k and T.~Madaras.
\newblock On local structure of 1-planar graphs of minimum degree 5 and girth
  4.
\newblock {\em Discuss. Math. Graph Theory}, 29(2):385--400, 2009.

\bibitem{HMS12}
D.~Hud{\'a}k, T.~Madaras, and Y.~Suzuki.
\newblock On properties of maximal 1-planar graphs.
\newblock {\em Discuss. Math. Graph Theory}, 32(4):737--747, 2012.

\bibitem{HudakS12}
D.~Hud{\'{a}}k and P.~{\v S}ugerek.
\newblock Light edges in $1$-planar graphs with prescribed minimum degree.
\newblock {\em Discuss. Math. Graph Theory}, 32(3):545--556, 2012.

\bibitem{Juenger04}
M.~J{\"{u}}nger and P.~Mutzel, editors.
\newblock {\em Graph Drawing Software}.
\newblock Springer, 2004.

\bibitem{Kainen90}
P.~C. Kainen.
\newblock The book thickness of a graph. {II}.
\newblock In {\em {CGTC} 1990}, number v. 5 in Congressus numerantium, pages
  127--132. Utilitas Mathematica, 1990.

\bibitem{Karpov2014}
D.~V. Karpov.
\newblock An upper bound on the number of edges in an almost planar bipartite
  graph.
\newblock {\em J. Math. Sci.}, 196(6):737--746, 2014.

\bibitem{kw-dgmm-01}
M.~Kaufmann and D.~Wagner, editors.
\newblock {\em Drawing Graphs, Methods and Models}, volume 2025 of {\em LNCS}.
  Springer, 2001.

\bibitem{Kedlaya1996238}
K.~S. Kedlaya.
\newblock Outerplanar partitions of planar graphs.
\newblock {\em J. Combin. Theory Ser. B}, 67(2):238 -- 248, 1996.

\bibitem{KM13}
V.~P. Korzhik and B.~Mohar.
\newblock Minimal obstructions for 1-immersions and hardness of 1-planarity
  testing.
\newblock {\em J. Graph Theory}, 72(1):30--71, 2013.

\bibitem{KralS10}
D.~Kr{\'{a}}l' and L.~Stacho.
\newblock Coloring plane graphs with independent crossings.
\newblock {\em J. Graph Theory}, 64(3):184--205, 2010.

\bibitem{LenhartLM17}
W.~J. Lenhart, G.~Liotta, and F.~Montecchiani.
\newblock On partitioning the edges of 1-plane graphs.
\newblock {\em Theor. Comput. Sci.}, 662:59--65, 2017.

\bibitem{Lewis16}
R.~Lewis.
\newblock {\em A Guide to Graph Colouring: Algorithms and Applications}.
\newblock Springer, 2016.

\bibitem{Liotta14}
G.~Liotta.
\newblock Graph drawing beyond planarity: some results and open problems.
\newblock In {\em {ICTCS} 2014.}, volume 1231 of {\em {CEUR} Workshop
  Proceedings}, pages 3--8. CEUR-WS.org, 2014.

\bibitem{LiottaM16}
G.~Liotta and F.~Montecchiani.
\newblock {L}-visibility drawings of {IC}-planar graphs.
\newblock {\em Inf. Process. Lett.}, 116(3):217--222, 2016.

\bibitem{Malitz1994}
S.~Malitz.
\newblock Graphs with {$E$} edges have pagenumber {$O(\sqrt{E})$}.
\newblock {\em J. Algorithms}, 17(1):71 -- 84, 1994.

\bibitem{Marx2004}
D.~Marx.
\newblock Graph colouring problems and their applications in scheduling.
\newblock {\em Periodica Polytechnica Ser. El. Eng.}, 48(1):11--16, 2004.

\bibitem{McKenzieO10}
T.~McKenzie and S.~Overbay.
\newblock Book embeddings and zero divisors.
\newblock {\em Ars Comb.}, 95, 2010.

\bibitem{Mohar97}
B.~Mohar.
\newblock Circle packings of maps in polynomial time.
\newblock {\em Eur. J. Comb.}, 18(7):785--805, 1997.

\bibitem{DBLP:books/daglib/0030491}
J.~Ne{\v s}et{\v r}il and P.~O. de~Mendez.
\newblock {\em Sparsity - Graphs, Structures, and Algorithms}, volume~28 of
  {\em Algorithms and combinatorics}.
\newblock Springer, 2012.

\bibitem{NesetrilMW12}
J.~Ne{\v s}et{\v r}il, P.~O. de~Mendez, and D.~R. Wood.
\newblock Characterisations and examples of graph classes with bounded
  expansion.
\newblock {\em Eur. J. Comb.}, 33(3):350--373, 2012.

\bibitem{NishizekiB79}
T.~Nishizeki and I.~Baybars.
\newblock Lower bounds on the cardinality of the maximum matchings of planar
  graphs.
\newblock {\em Discr. Math.}, 28(3):255--267, 1979.

\bibitem{NR04}
T.~Nishizeki and M.~S. Rahman.
\newblock {\em Planar graph drawing}, volume~12 of {\em Lecture notes series on
  computing}.
\newblock World Scientific, 2004.

\bibitem{ov-grild-78}
R.~H. J.~M. Otten and J.~G.~V. Wijk.
\newblock Graph representations in interactive layout design.
\newblock In {\em {IEEE ISCSS}}, pages 914--918. IEEE, 1978.

\bibitem{PachT97}
J.~Pach and G.~T{\'o}th.
\newblock Graphs drawn with few crossings per edge.
\newblock {\em Combinatorica}, 17(3):427--439, 1997.

\bibitem{Purchase97}
H.~C. Purchase.
\newblock Which aesthetic has the greatest effect on human understanding?
\newblock In {\em {GD} 1997}, volume 1353 of {\em LNCS}, pages 248--261.
  Springer, 1997.

\bibitem{Purchase00}
H.~C. Purchase.
\newblock Effective information visualisation: a study of graph drawing
  aesthetics and algorithms.
\newblock {\em Interacting with Computers}, 13(2):147--162, 2000.

\bibitem{PurchaseCA02}
H.~C. Purchase, D.~A. Carrington, and J.~Allder.
\newblock Empirical evaluation of aesthetics-based graph layout.
\newblock {\em Empirical Software Engineering}, 7(3):233--255, 2002.

\bibitem{R65}
G.~Ringel.
\newblock Ein {S}echsfarbenproblem auf der kugel.
\newblock {\em Abhandlungen aus dem Mathematischen Seminar der Universitaet
  Hamburg}, 29(1--2):107--117, 1965.

\bibitem{RosenstiehlT86}
P.~Rosenstiehl and R.~E. Tarjan.
\newblock Rectilinear planar layouts and bipolar orientations of planar graphs.
\newblock {\em Discr. {\&} Comput. Geom.}, 1:343--353, 1986.

\bibitem{Shermer96}
T.~C. Shermer.
\newblock On rectangle visibility graphs. {III.} {E}xternal visibility and
  complexity.
\newblock In {\em {CCCG} 1996}, pages 234--239. Carleton University Press,
  1996.

\bibitem{Sugiyama02}
K.~Sugiyama.
\newblock {\em Graph Drawing and Applications for Software and Knowledge
  Engineers}, volume~11 of {\em Series on Software Engineering and Knowledge
  Engineering}.
\newblock World Scientific, 2002.

\bibitem{SukW15}
A.~Suk and B.~Walczak.
\newblock New bounds on the maximum number of edges in $k$-quasi-planar graphs.
\newblock {\em Comput. Geom.}, 50:24--33, 2015.

\bibitem{S10b}
Y.~Suzuki.
\newblock Optimal 1-planar graphs which triangulate other surfaces.
\newblock {\em Discrete Math.}, 310(1):6--11, 2010.

\bibitem{S10}
Y.~Suzuki.
\newblock Re-embeddings of maximum 1-planar graphs.
\newblock {\em {SIAM} J. Discrete Math.}, 24(4):1527--1540, 2010.

\bibitem{Tamassia99}
R.~Tamassia.
\newblock Advances in the theory and practice of graph drawing.
\newblock {\em Theor. Comput. Sci.}, 217(2):235--254, 1999.

\bibitem{2013gd}
R.~Tamassia, editor.
\newblock {\em Handbook on Graph Drawing and Visualization}.
\newblock Chapman \& Hall/CRC, 2013.

\bibitem{TamassiaTollis86}
R.~Tamassia and I.~G. Tollis.
\newblock A unified approach to visibility representations of planar graphs.
\newblock {\em Discr. \& Comput. Geom.}, 1(1):321--341, 1986.

\bibitem{t-prg-84}
C.~Thomassen.
\newblock Plane representations of graphs.
\newblock In {\em Progress in Graph Theory}, pages 43--69. AP, 1984.

\bibitem{Tho88}
C.~Thomassen.
\newblock Rectilinear drawings of graphs.
\newblock {\em J. Graph Theory}, 12(3):335--341, 1988.

\bibitem{Thorup98}
M.~Thorup.
\newblock Map graphs in polynomial time.
\newblock In {\em {FOCS} 1998}, pages 396--405. {IEEE}, 1998.

\bibitem{Vizing64}
V.~G. Vizing.
\newblock On an estimate of the chromatic class of a {$p$}-graph.
\newblock {\em Diskret. Analiz No.}, 3:25--30, 1964.

\bibitem{WL08}
W.~Wang and K.-W. Lih.
\newblock Coupled choosability of plane graphs.
\newblock {\em J. Graph Theory}, 58(1):27--44, 2008.

\bibitem{Wattenberg2002}
M.~Wattenberg.
\newblock Arc diagrams: visualizing structure in strings.
\newblock In {\em IEEE {INFOVIS} 2002.}, pages 110--116. IEEE, 2002.

\bibitem{whitney32}
H.~Whitney.
\newblock Congruent graphs and the connectivity of graphs.
\newblock {\em Am. J. Math.}, 54(1):150--168, 1932.

\bibitem{Wismath85}
S.~K. Wismath.
\newblock Characterizing bar line-of-sight graphs.
\newblock In {\em {SoCG} 1985}, pages 147--152. {ACM}, 1985.

\bibitem{Wood2002}
D.~R. Wood.
\newblock Bounded degree book embeddings and three-dimensional orthogonal graph
  drawing.
\newblock In {\em {GD} 2001}, volume 2265 of {\em LNCS}, pages 312--327.
  Springer, 2002.

\bibitem{Yannakakis1989}
M.~Yannakakis.
\newblock Embedding planar graphs in four pages.
\newblock {\em J. Comput. Syst. Sci.}, 38(1):36 -- 67, 1989.

\bibitem{Zhang2014}
X.~Zhang.
\newblock Drawing complete multipartite graphs on the plane with restrictions
  on crossings.
\newblock {\em Acta Math. Sin. English Series}, 30(12):2045--2053, 2014.

\bibitem{Zhang14d}
X.~Zhang.
\newblock The edge chromatic number of outer-1-planar graphs.
\newblock {\em CoRR}, abs/1405.3183, 2014.

\bibitem{ZhangHL15}
X.~Zhang, J.~Hou, and G.~Liu.
\newblock On total colorings of 1-planar graphs.
\newblock {\em J. Comb. Optim.}, 30(1):160--173, 2015.

\bibitem{ZhangL2012}
X.~Zhang and G.~Liu.
\newblock On edge colorings of 1-planar graphs without adjacent triangles.
\newblock {\em Inf. Process. Lett.}, 112(4):138 -- 142, 2012.

\bibitem{ZhangL12}
X.~Zhang and G.~Liu.
\newblock On edge colorings of 1-planar graphs without chordal 5-cycles.
\newblock {\em Ars Comb.}, 104:431--436, 2012.

\bibitem{Zhang2013}
X.~Zhang and G.~Liu.
\newblock The structure of plane graphs with independent crossings and its
  applications to coloring problems.
\newblock {\em Open Math.}, 11(2):308--321, 2013.

\bibitem{Zhang12}
X.~Zhang, G.~Liu, and J.-L. Wu.
\newblock Edge covering pseudo-outerplanar graphs with forests.
\newblock {\em Discrete Math.}, 312(18):2788--2799, 2012.

\bibitem{ZWL12}
X.~Zhang, J.~Wu, and G.~Liu.
\newblock List edge and list total coloring of 1-planar graphs.
\newblock {\em Front. Math. China}, 7(5):1005--1018, 2012.

\bibitem{ZW11}
X.~Zhang and J.-L. Wu.
\newblock On edge colorings of 1-planar graphs.
\newblock {\em Inf. Process. Lett.}, 111(3):124--128, 2011.

\end{thebibliography}
\bibliographystyle{abbrv}
\end{document}